\documentclass[conference]{IEEEtran}
\IEEEoverridecommandlockouts
\usepackage{cite}
\usepackage{amsmath,amssymb,amsfonts}
\usepackage{graphicx}
\usepackage{textcomp}
\usepackage{xcolor}
\usepackage[hyphens]{url}
\usepackage{fancyhdr}
\usepackage[bookmarks=true,breaklinks=true,letterpaper=true,colorlinks,citecolor=blue,linkcolor=blue,urlcolor=blue]{hyperref}

\usepackage{braket}
\usepackage{float}
\usepackage{tikz}
\usepackage{siunitx}
\usepackage{listings}
\usepackage{subfigure}
\usepackage{multirow}
\usepackage[normalem]{ulem}
\usepackage[linesnumbered,ruled,vlined]{algorithm2e}
\usepackage{algpseudocode}
\usepackage{booktabs}
\usepackage{natbib}

\fancypagestyle{firstpage}{
  \fancyhf{}
  
  \fancyfoot[C]{\thepage}
}

\SetAlCapFnt{\small} 
\SetAlCapNameFnt{\small} 
\SetAlFnt{\footnotesize}

\usepackage{soul} % \st{xxx} = delete
\usepackage[affil-it]{authblk}

\title{LATTE: A Decoding Architecture for Quantum Computing with Temporal and Spatial Scalability} 
\author[1]{Kai Zhang\textsuperscript{$\dagger$}\thanks{$\dagger$ Equal contribution.}}
\author[2]{Jubo Xu\textsuperscript{$\dagger$}}
\author[3]{Fang Zhang\textsuperscript{\S}}
\author[3]{Linghang Kong}
\author[1,3]{Zhengfeng Ji\textsuperscript{\S}}
\author[1]{Jianxin Chen\textsuperscript{\S}\thanks{$\S$ Corresponding author: fangzhang@iqubit.org, \{jizhengfeng,\ chenjianxin\}@tsinghua.edu.cn}}

\affil[1]{Department of Computer Science and Technology, Tsinghua University, Beijing, China}
\affil[2]{Independent Researcher}
\affil[3]{Zhongguancun Laboratory, Beijing, China}
\begin{document}
\maketitle
\thispagestyle{firstpage}
\pagestyle{plain}

%%%%%% -- PAPER CONTENT STARTS-- %%%%%%%%

\begin{abstract}
Quantum error correction allows inherently noisy quantum devices to emulate an ideal quantum computer with reasonable resource overhead. As a crucial component in developing fault-tolerant quantum computers, decoding architectures --- designed to effectively manage a large number of physical qubits to ensure reliable computation --- have received significant attention recently, particularly with the rapid advancements in quantum hardware.
  
In this paper, we introduce LATTE, a FPGA-CPU hybrid decoding architecture aiming to address the key requirements of scaling up, especially in lattice surgery quantum computation --- \underline{L}atency, \underline{A}ccuracy, \underline{T}hroughput and \underline{T}ransmission Bandwidth, in an \underline{E}clectic manner. LATTE follows a hierarchical design: (1) A fully streaming and asynchronous block decoding system on CPU to enable parallelization both temporally and spatially. (2) A super-light yet accurate neural local decoding unit integrated with quantum control hardware on FPGA, which remains \emph{transparent} to the block decoding system, effectively reducing transmission bandwidth and accelerating the decoding process.

As a decoding architecture, LATTE is compatible with various base decoders, delivering accuracy on par with the base decoder while achieving real-time decoding throughput and significantly reducing both bandwidth requirements and computational resources, enabling a level of scalability far beyond previous approaches. Under circuit-level noise $p=0.001$, LATTE achieves over $\mathbf{90\%}$ reduction in transmission bandwidth and a $\mathbf{6.4\times}$ speedup on average in single-block decoding. In the \emph{streaming decoding} scenario for logical quantum operations: (1) LATTE achieves constant and low latency ($\mathbf{16\times}$-$\mathbf{20\times}$ speedup over existing streaming decoding implementations) in arbitrarily long quantum memory experiments, with near-optimal computational resources --- merely $\mathbf{2}$ threads are sufficient for decoding the surface code with distance up to $17$. (2) LATTE minimizes latency in multi-patch measurement experiments through highly parallelized decoding operations. These combined efforts ensure sufficient scalability for large-scale fault-tolerant quantum computing.

\end{abstract}

\section{Introduction}
\label{sec:introduction}
Large-scale quantum computers promise computational power far beyond that of traditional computers~\cite{lloyd1996universal,shor1999polynomial,grover1996fast,peruzzo2014variational}. However, physical qubits are highly susceptible to noise and require correction to perform quantum computations reliably. Quantum error correction (QEC), the mechanism for correcting these errors, generates a continuous stream of error data, known as syndromes, which must be processed by decoders. To achieve a sufficiently small logical error, the code distance must be large enough, necessitating the use of a large number of physical qubits to encode a single logical qubit. Consequently, QEC must process syndrome data on a vast number of physical qubits efficiently. To this end, the decoding architecture that orchestrates massive resources, whether quantum or classical, becomes crucial for the development of fault-tolerant quantum computing (FTQC). There are several key aspects to consider in a decoding architecture:

\begin{itemize}
    \item {\emph{Accuracy}}: High decoding accuracy is essential for maintaining the integrity of quantum information and ensuring the overall reliability and performance of the quantum computing system.
    \item {\emph{Transmission Bandwidth}}: A vast amount of syndrome data needs to be transmitted from the quantum chip to the decoding unit during FTQC. Therefore, the decoding architecture must support sufficient bandwidth to keep up with the QEC cycles without causing a transmission backlog~\cite{delfosse2020hierarchical}.
    \item {\emph{Throughput}}: The well-recognized backlog problem also requires extensive syndrome information to be processed in real-time~\cite{terhal2015quantum}, potentially as quickly as $1 \mu s$ per QEC round in superconducting quantum computers, to match the syndrome extraction rate and avoid memory overflow or exponentially slowing down the computation.
    \item {\emph{Latency}}: When throughput and bandwidth requirements are met, there can still be a latency between the last measurement round and decoding feedback that is constant over time. For non-Clifford gate implementation in lattice surgery~\cite{litinski2019game}, reducing the latency improves the fidelity.
\end{itemize}

These aspects have been identified through an extensive body of research on decoding architectures for QEC. While much of this research has primarily focused on addressing specific challenges, there has been a recent shift towards a more holistic approach:

To achieve better \emph{accuracy}, different decoding algorithms have been proposed~\cite{criger2018multi,wu2023fusion,higgott2025sparse, wu2024hypergraph,varbanov2023neural, egorov2023end,bausch2024learning}, while mostly requiring complex graph data structures and large memory cost for decoding in lattice surgery. This indicates that we still need general-purpose computing units such as CPUs or GPUs to run high-accuracy decoding algorithms, supporting near-term demonstration. 

The first paper to highlight the \emph{transmission bandwidth} problem was~\cite{delfosse2020hierarchical}, although the magnitude of the problem was somewhat exaggerated as the paper did not take into account the fact that even the raw syndromes are quite sparse and have much less than 1 bit of entropy per ancilla qubit~\cite{das2022afs}. Many other predecoders~\cite{Holmes2020NISQBQ, ravi2023better, Vittal2023AstreaAQ, Smith_2023, alavisamani2024promatch, meinerz2022scalable, Chamberland2022TechniquesFC, caune2023belief} are proposed later on to reduce the syndrome density and bandwidth overhead from different perspectives. Some predecoders like~\cite{delfosse2020hierarchical, ravi2023better} were applied too conservatively, resulting in little accuracy loss but also virtually no bandwidth reduction. On the other hand, more aggressive predecoders like~\cite{Smith_2023, Vittal2023AstreaAQ, alavisamani2024promatch, Holmes2020NISQBQ} requires complex human-design, but performing well only at low physical error rates or small decode distances. Using neural network (NN) or belief propagation (BP) as predecoders has been tested in~\cite{meinerz2022scalable, wang2023transformer, Chamberland2022TechniquesFC, caune2023belief}. Although some achieved better accuracy for limited code distances, they relied on large computation resources and were not scalable for control hardware integration. Moreover, previous NN or BP predecoders produce latency much higher than $1\mu s$ per round, resulting the backlog even in the hardware level, making them unsuitable for real scenarios in lattice surgery. Therefore, existing predecoders are still far from perfect, calling for more accurate and scalable local decoding designs.

To avoid syndrome backlog and maintain high \emph{throughput}, parallel sliding window decoding algorithms~\cite{Tan2022ScalableSD,skoric2023parallel,bombin2023modular, lin2024spatially} have been proposed. However, during practical decoding processes where syndromes arrive continuously, achieving temporal and spatial parallel decoding with better scheduling strategy and optimal resources remains challenging and will improve the system's scalability.

Works like~\cite{das2022lilliput, das2022afs, barber2025real, alavisamani2024promatch} opt to implement the decoder completely on FPGAs or ASICs to demonstrate extremely low \emph{latency} close to $1 \mu s$ on a single $d \times d \times d$ block. However, the strict time constraint limits their scalability for larger error rates or code distances. Meanwhile, in practical decoding scenarios, as long as the throughput is sufficient, latency below $1 \mu s$ is not strictly required, demonstrated by~\cite{wu2023fusion,acharya2025quantum}. It is more useful to optimize the latency under more reasonable error rates and code distances. Considering the graph deformation and frequent feedback during lattice surgery, employing global decoding on general-purpose computing units remains the most scalable option, at least for now.

\textbf{Key Contributions.} To address challenges of existing decoding approaches and build up a decoding system with temporal and spatial scalability for practical deployment, we propose and implement a streaming and distributed decoding architecture with high scalability and excellent performance. In summary, the contributions of our work are as follows: 

\begin{itemize}
  \item \textit{{Fully Streaming and Distributed Decoding Architecture}}: We propose a modular and distributed FPGA-CPU integrated decoding architecture for scalability, combining syndrome pre-processing, decoding and feedback into a fully streaming pipeline. The design replicates the streaming behavior of real-world QEC process, achieves near-optimal resource efficiency while natively supporting temporally and spatially flexible volumetric lattice surgery computations.
  \item \textit{{Asynchronous Block Decoding}}: We extend sliding window decoder into an asynchronous decoding system, enabling parallelization across both temporal and spatial dimensions, allowing decoding feedback with constant and low latency.
  \item \textit{Temporally-Spatially Local Neural Decoding Unit}: We design and implement an extremely lightweight neural network on FPGA. Coupled with a 3-stage streaming pipeline and a $(7,7, 7)$ receptive field for temporally-spatially local decoding, it ensures sufficient accuracy, highly reduced transmission bandwidth and low decoding latency.
\end{itemize}

Extensive benchmarking demonstrated that our architecture offers strong temporal and spatial scalability when integrated with quantum control devices, even in the presence of noise levels typical of current-generation hardware --- a level of scalability not achieved by previous approaches.

\begin{figure}[htb]
    \centering
    \includegraphics[width=\linewidth]{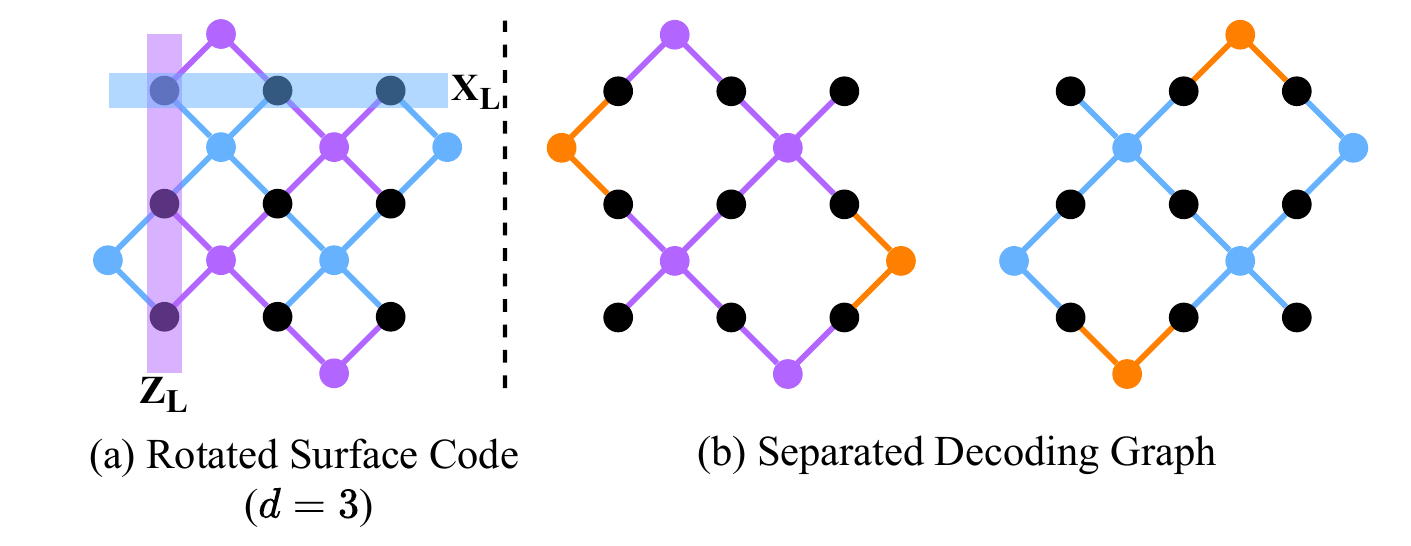}
    \caption{(a) An illustration of the $d = 3$ rotated surface code with logical Z/X operators. (b) Separated decoding graph of Z and X stabilizers. The black, purple, blue and orange vertices represent (respectively) data qubits, Z stabilizers, X stabilizers, and virtual vertices assisting in decoding the errors of boundary data qubits.}
    \label{fig:surface code and separated decoding graph}
\end{figure}

\section{Background and Motivation}
\label{sec:background}
In this section, we will provide background on quantum error correction and present our motivation for designing the architecture.

\subsection{Surface Code}
\label{sec:surface_code}

Throughout this work, we will focus on the superconducting quantum chip, as it is one of the leading platforms. The superconducting quantum computing platform also presents greater challenges for the decoding architecture due to its rapid operations, serving as a stress test for the system. The mainstream error correction scheme for superconducting qubits is the surface code~\cite{Fowler_2012} illustrated in Figure~\ref{fig:surface code and separated decoding graph}\textcolor{blue}{(a)}, which encodes one logical qubit into a $d\times d$ array of physical qubits with code distance $d$. The threshold theorem~\cite{aharonov1997fault, kitaev1997quantum, Knill1998ResilientQC} guarantees that the logical error rate can be exponentially suppressed as the code distance increases, provided that the physical error rate $p$ is below the fault-tolerant threshold.
\begin{figure}[htb]
    \centering
    \includegraphics[width=\linewidth]{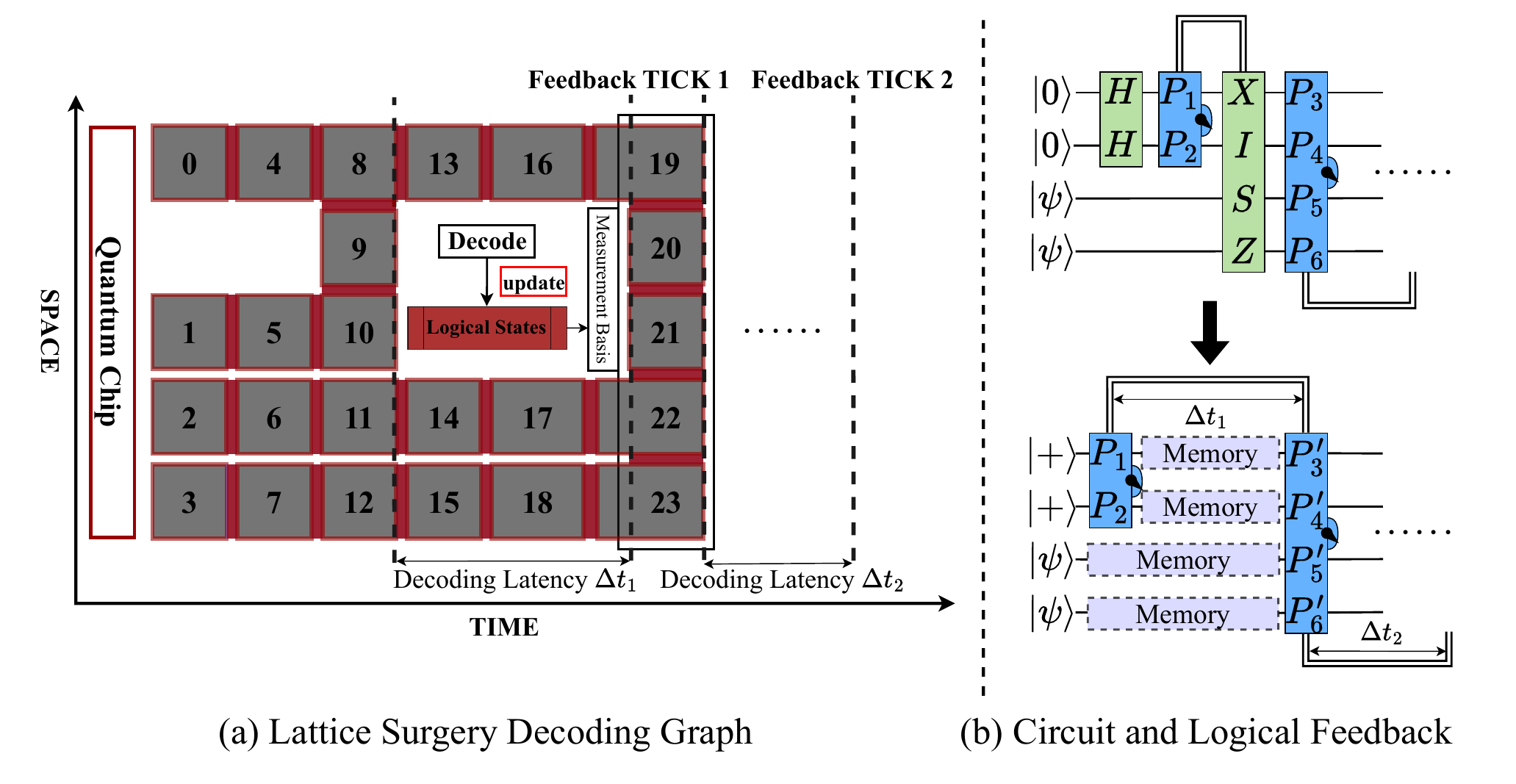}
    \caption{(a) Temporal and spatial decoding graph example during lattice surgery, where the decoding tasks can be 
    % distributed both temporally and spatially, and 
    partitioned by LATTE according to their temporal order as $0$ to $23$. (b) Circuit and logical feedback. 
    % The circuit during lattice surgery~\cite{litinski2019game} can be compiled from~\cite{watkins2024high, liu2023substrate, tan2024sat} as a pipelined,  
    The logical results of decoding are used to determine the logical measurement bases at next feedback TICK during the entire quantum program.}
    \label{fig:space and time decoding graph}
\end{figure}

Errors occurring on data qubits will flip the adjacent stabilizers as Z errors anti-commute with $X$ stabilizers and X errors anti-commute with $Z$ stabilizers. Additionally, the readout process of stabilizer ancilla qubits also introduces errors, resulting in measurement errors as well as more complicated errors that can be propagated back onto data qubits. To localize the influence of errors for decoding input, we usually XOR the syndrome outcomes of adjacent measurement rounds to form \emph{detectors}~\cite{Gidney2021StimAF}. As described in~\cite{wu2023fusion,higgott2025sparse,Dennis_2002}, the decoding problem can be modeled using the \emph{decoding graph} $(V, E)$.  A detector result corresponds to a $v \in V$, while an $e \in E$ is an independent error flipping adjacent detectors. To be more specific, the circuit-level noise model~\cite{Dennis_2002} includes 4 types of errors: data qubit errors $\{I,X,Y,Z\}$, measurement errors $\{M\}$, diagonal errors $\{D\}$ and hook errors $\{H\}$. Each type of error will flip detectors in a certain pattern. 

Quantum error correction involves identifying the most likely errors through decoding algorithms such as minimum weight perfect matching (MWPM)~\cite{Edmonds1973MatchingET, Kolmogorov2009BlossomVA} or union-find (UF)~\cite{delfosse2021almost}, updating the logical states using Pauli Frame~\cite{Riesebos2017PauliFF}, thus realizing fault tolerant quantum logical operations.

\subsection{Quantum Error Pattern}
\label{sec:quantum error pattern}
Due to the complexity of quantum error patterns, hand-designed greedy matching methods~\cite{delfosse2020hierarchical,Holmes2020NISQBQ,ravi2023better,Smith_2023,alavisamani2024promatch,Vittal2023AstreaAQ} often struggle to handle intricate situations, leading to a loss in accuracy and efficiency. However, errors occurring on the surface code exhibit strong characteristics, 
% compared to classical codes
indicating the potential for more efficient error correction.

\textbf{Errors Are Local and Sparse.} Most errors are sparse and can be locally identified, 
which has motivated a line of research focused on predecoders~\cite{delfosse2020hierarchical,ravi2023better,Holmes2020NISQBQ,alavisamani2024promatch, Vittal2023AstreaAQ,Smith_2023}. One advantage of syndrome pre-processing is that it can reduce the transmission bandwidth and speed up the overall decoding. To figure out a reasonable ``receptive field'', we estimate the error chain length distribution at current noise level $p \sim 10^{-3}$. Numerical experiments indicate that over 90\% of error chains still have a length $\leq2$, even at $d = 21$. Therefore, we should design local decoding with input similar to this size to balance pre-processing complexity and accuracy. 

Existing pre-decoding algorithms must be executed sequentially or with a global check, which significantly reduces their decoding efficiency. However, the inherent sparsity of errors suggests that local syndrome information can be processed in parallel without substantially compromising overall performance.

\textbf{Correlated Errors Are Common.} In the surface code, a physical error with both X and Z components, such as a single-qubit Y error or a two-qubit XZ error, can flip up to four detectors. Therefore the decoding graph introduced in $\S$~\ref{sec:surface_code} is, in fact, a hypergraph.
% where ancilla qubit measures the syndromes of X and Z stabilizers, 
Traditionally, decoding algorithms separate the Z and X decoding graphs as depicted in Figure~\ref{fig:surface code and separated decoding graph}\textcolor{blue}{(b)}, thereby bypassing the challenge of matching on Z and X sub-graphs together. In this separation, a physical error corresponding to a hyperedge is decomposed into one edge on the Z sub-graph and one on the X sub-graph. Decoding each sub-graph separately is essentially making an approximation that, after this decomposition, errors on the Z sub-graph and the X sub-graph still occur independently, which is far from accurate. Due to the depolarizing noise, while the \emph{a priori} probability of a post-decomposition X or Z error is $O(p)$ as expected, the conditional probability of a Z error occurring given that an X error has already occurred is as high as $1/2$~\cite{fowler2013optimal}, and vice versa. In other words, there is a \emph{correlation} between the two types of errors.

Some decoding schemes can take error correlation into account~\cite{breuckmann2018scalable,caune2023belief, bausch2024learning, wu2024hypergraph} and achieve higher accuracy than decoding X or Z syndrome independently. Therefore, we aim to leverage the properties of hypergraph decoding to achieve more precise error identification. 

\subsection{Decoding Tasks in Time and Space}
\label{sec:temporal-spatial decoding}
A well-known fault-tolerant quantum computing scheme based on the surface code is lattice surgery~\cite{litinski2019game}, which encodes physical qubits on a lattice into logical qubits. All logical operations needed for universal quantum computation can be carried out through local operations on the physical qubits, with fault tolerance ensured by continuous syndrome extraction and error decoding. 

In lattice surgery, decoding tasks are distributed both temporally and spatially as illustrated in Figure~\ref{fig:space and time decoding graph}. The decoding graph in Figure~\ref{fig:space and time decoding graph}\textcolor{blue}{(a)} corresponds to the quantum circuit in Figure~\ref{fig:space and time decoding graph}\textcolor{blue}{(b)}, consisting of a series of measurement-based gates, which are necessary for the implementation of non-Clifford gates~\cite{litinski2019game}. The Pauli Frame~\cite{Riesebos2017PauliFF} on Clifford gates $C$ and Pauli measurements $M_P$ shows:
\[C \circ M_{P'} = M_P \circ C\]
where $P' = C^\dagger P C$ denotes the new measurement basis. Thus, the decoded joint-Pauli measurement outcome of $M_{P_1 P_2}$ controls not an explicit gate $X \otimes I \otimes S \otimes Z$, but the next measurement basis $P'_3 \otimes P'_4 \otimes P'_5 \otimes P'_6$:
\[P'_3 \otimes P'_4 \otimes P'_5 \otimes P'_6 = P_3 \otimes P_4 \otimes P_5 \otimes P_6, \text{if } M_{P_1 P_2} = 0\]
\[\begin{aligned}
    P'_3 \otimes P'_4 \otimes P'_5 \otimes P'_6 &= (X^\dagger P_3 X) \otimes (P_4) \otimes (S^\dagger P_5 S) \otimes (Z^\dagger P_6 Z), \\& \text{if } M_{P_1 P_2} = 1
\end{aligned}
\]

For circuits with many non-Clifford gates, the tracking of Pauli measurement bases thus requires frequent logical feedback from the decoder, forming a \emph{dependency chain}~\cite{erhard2021entangling, litinski2019game, bombin2023modular}. Due to the unavoidable classical decoding time, e.g. $\Delta t_1$ and $\Delta t_2$ shown in Figure~\ref{fig:space and time decoding graph}\textcolor{blue}{(b)}, some logical patches have to be kept idle until the outcome of the decoder and the execution of the next Pauli measurement. Those patches must be protected by continued syndrome extraction cycles.

Importantly, the Pauli frame technique~\cite{Riesebos2017PauliFF} allows a Pauli measurement to be \emph{executed} as long as the basis is known, without decoding all previous syndrome. Therefore a decoding latency, even if $> t_\text{round} \approx 1 \mu s$, does not prevent continuous syndrome extraction. In \textbf{Feedback TICK 1}, syndromes generated due to the idling in $\Delta t_1$ are only used to decode the measurement \emph{result} of $M_{P'_3P'_4P'_5P'_6}$, and the same applies to the next feedback TICK.

% Therefore, the tracking of Pauli frame in lattice surgery requires frequent logical feedback from the decoder and forms a \emph{dependency chain}~\cite{erhard2021entangling, litinski2019game, bombin2023modular}. Due to the unavoidable classical decoding time, e.g. $T_1$ and $T_2$ shown in Figure~\ref{fig:space and time decoding graph}\textcolor{blue}{(b)}, logical patches have to be kept in the quantum memory with continuous syndrome measurement $1\mu s$ per round before the outcome of the decoder and the execution of the next Pauli measurement. In \textbf{Feedback TICK 1}, the syndrome generated due to the idling in $T_1$ will still be used, but to decode the measurement result of $M_{P'_3P'_4P'_5P'_6}$, and the same applies to the next feedback TICK.

Suppose the logical error rate per round is $\epsilon$, a decoding latency $T$ will cause an increase on the total logical error rate (LER)~\cite{bausch2024learning}:
\[\Delta LER \approx \frac{1}{2} (1 - (1 - 2\epsilon)^{n}) \approx n \epsilon, n = \lceil \frac{T}{1\mu s}\rceil\]

Therefore, a constant decoding latency only leads to a near-linear degradation in logical fidelity, as $n\epsilon = O(np^{\left\lceil \frac{d}{2} \right\rceil})$ is still exponentially low. In contrast, if the latency increases with the decoding graph size --- scaling with the temporal-spatial volume in lattice surgery --- then decoding tasks will accumulate exponentially over time, ultimately creating a backlog that destroys the FTQC pipeline.

However, compilation of an arbitrary circuit into lattice surgery primitives remains complicated~\cite{watkins2024high, liu2023substrate, tan2024sat}, and decoding a compiled lattice surgery circuit would involve even more engineering challenges. To simplify and benchmark the process for the decoding architecture, lattice surgery can be modeled as the following two processes:

\textbf{Arbitrarily Long Quantum Memory.} Quantum memory is one of the most common components in lattice surgery quantum computation, as all logical qubits must be kept in memory before measurement shown in Figure~\ref{fig:space and time decoding graph}\textcolor{blue}{(b)}. The realization of arbitrarily long quantum memory is also a significant milestone in FTQC, making memory experiments a primary method for validating or benchmarking the reliability of quantum memory and decoding algorithms. In the standard memory experiment, the quantum logical state encoded in a $d \times d$ surface code is usually initialized to $\ket{0_L}$. After $d$ rounds of syndrome measurement, the decoding unit performs the decoding algorithm, identifies the errors and performs the corrections. Finally, a logical measurement is conducted. If the result is still $\ket{0_L}$, the decoding is successful. Otherwise, i.e.\ if the result is $\ket{1_L}$, it will be regarded as a failure, indicating that one logical error has occurred. Achieving arbitrarily long quantum memory places greater demands on decoders compared to the $d\times d\times d$ case, requiring not only high accuracy but also the ability to process continuously measured syndromes in a streaming fashion without accumulating a backlog over time. This necessitates a decoder with high average decoding speed, or throughput~\cite{Tan2022ScalableSD, Zhang2023ACA}.

\textbf{Highly Scalable Multi-Patch Measurement.} 
In contrast to quantum memory, logical operations in lattice surgery pose challenges especially in the space  dimension. 
Most logical operations require joint decoding across multiple logical qubits to get the logical measurement results. During the multi-patch measurement as shown in Figure~\ref{fig:space and time decoding graph}\textcolor{blue}{(b)}, supposing there are $k$ logical qubits patches and we want to measure $P_1 \otimes P_2 \otimes \cdots \otimes P_k$ in the Pauli basis ($P_i \in \{I,X,Y,Z\}$), the ancilla patches are usually initialized to $\ket{+_L}^{\otimes k - 1}$. After $d$ or more rounds of syndrome extraction and decoding, the $Z$ stabilizer product of ancilla patches is calculated to get the $P_1 \otimes P_2 \otimes \cdots \otimes P_k$ measurement result, which is also equal to the multiplication of $Z_1 \otimes P_1, Z_2 \otimes P_2, \cdots, Z_{k-1} \otimes P_{k-1}, (Z_1Z_2\cdots Z_{k-1}) \otimes P_k$.
This, in turn, demands high scalability of the overall decoding architecture to decode multi-patch simultaneously and efficiently.

\section{System Overview}
\begin{figure*}
\begin{center}
\includegraphics[width=\linewidth]{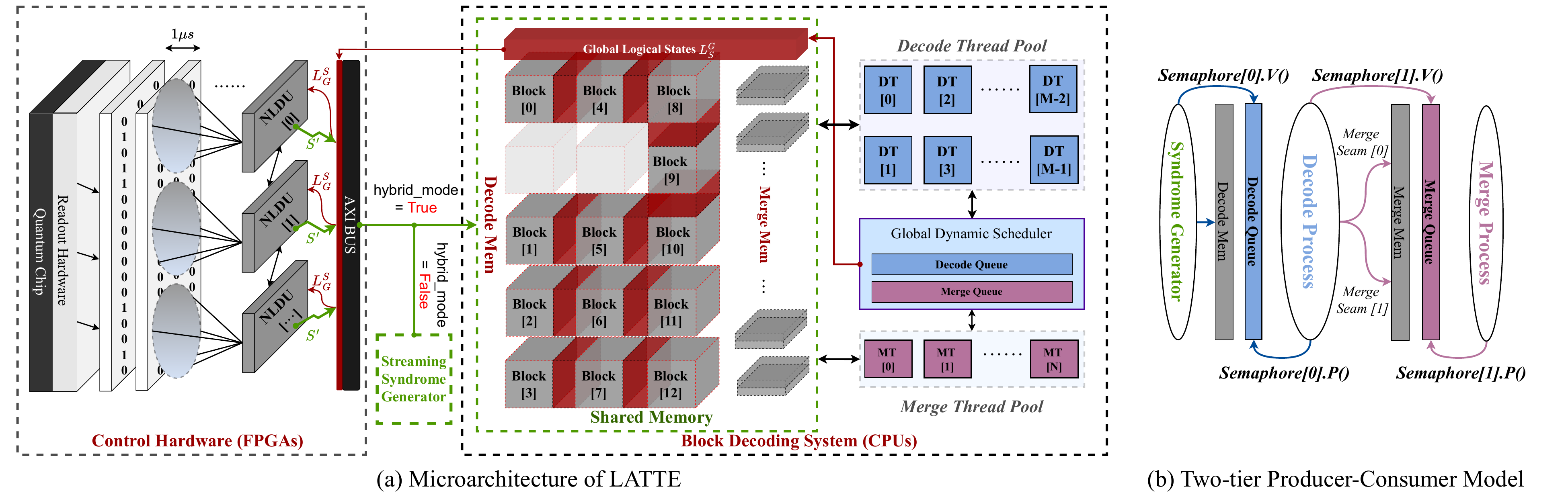}
\caption{Decoding architecture. (a) Microarchitecture of LATTE. The readout syndromes processed by the NLDU will output in a streaming manner to the BLDS, consistent with the syndrome transmission pattern of the \emph{Streaming Syndrome Generator}. All syndromes in the \emph{Shared Memory} are partitioned into different temporal or spatial blocks. Decoding and merging will be executed asynchronously to finish all decoding tasks, while the overall system is scheduled by the \emph{Global Dynamic Scheduler}. (b) Two-tier producer-consumer model to achieve asynchronous block decoding.}
\label{fig:decoding architecture pipeline}
\end{center}
\end{figure*}

To address the decoding requirements from a motivation-driven perspective, we begin with the parallel sliding window decoding algorithm and extend it to a fully streaming version with a dynamic scheduling strategy, addressing temporal and spatial tasks asynchronously. To reduce syndrome transmission bandwidth and minimize the overhead of CPU resources, we design a light-weight and accurate neural local decoding unit on FPGA and combine the overall system in a hierarchical scheme as in Figure~\ref{fig:decoding architecture pipeline}, achieving excellent streaming decoding performance in \emph{accuracy}, \emph{transmission bandwidth}, \emph{throughput} and \emph{latency} at the same time.

We will introduce our decoding architecture in a top-down manner. In $\S$~\ref{sec:block decoding}, we will provide an overview of the modules in the block decoding system, which form the system’s top layer, featuring its modularity, simplicity, and efficiency. In $\S$~\ref{sec:neural local decoding unit}, we will delve into the neural local decoding unit, designed for direct integration and naturally distributed in alignment with quantum control hardware. Since both $\S$~\ref{sec:block decoding} and $\S$~\ref{sec:neural local decoding unit} follow a fully streaming and distributed scheme, we develop them independently and integrate them ultimately in $\S$~\ref{sec:system implementation} into the full decoding architecture.

\section{Block Decoding System}
\label{sec:block decoding}
The block decoding system (BLDS) forms the CPU part of the hierarchical architecture. We describe the system through three submodules: the streaming syndrome generator, asynchronous block decoding and global dynamic scheduler. 

\subsection{Streaming Syndrome Generator} 

Most workflows of existing decoding approaches are overly static and simplified: The syndrome data are simulated and prepared in advance and then fed into the decoder. The decoder subsequently executes the algorithm, processing all syndromes in a single batch. This simulation process does not reflect real-world scenarios, where syndromes are generated continuously and must be processed in a streaming manner to avoid backlog.

To emulate the environment of streaming decoding, we utilizes a \emph{Streaming Syndrome Generator} as in Figure~\ref{fig:decoding architecture pipeline}\textcolor{blue}{(a)} to support the entire FTQC software pipeline. This generator mimics the syndrome readout process of FPGA, operating with a fixed time of $ 1 \mu s$, corresponding to superconducting quantum hardware.

The \emph{Streaming Syndrome Generator} allows great flexibility for developing the architecture. If \texttt{hybrid\_mode = \textcolor{red}{False}}, we can debug and verify our BLDS independently. However, if \texttt{hybrid\_mode = \textcolor{red}{True}}, in the actual FPGA-CPU hybrid deployment situation, we can directly replace the \emph{Streaming Syndrome Generator} with the quantum control hardware on FPGA, enabling seamless integration.

\subsection{Asynchronous Block Decoding} 

Parallel sliding window decoders focus on parallelizing decoding primarily along the time dimension. They divide the global decoding graph into overlapping time windows, which are then assigned to different threads for parallel \emph{decoding}, resolving boundary conflicts through \emph{merging}. The inner-window decoding algorithms, such as MWPM~\cite{higgott2022pymatching, higgott2025sparse} or Union-Find~\cite{delfosse2021almost,liyanage2024fpga}, can be flexibly chosen based on practical requirements, offering inherent modularity.

Although previous works have discussed how to partition windows for parallel decoding, there has been no system-level design on how to efficiently allocate computational resources to ensure that decoding and merging tasks are completed as parallel as possible. In complex scenarios during lattice surgery~\cite{litinski2019game}, where decoding tasks are temporally and spatially distributed as Figure~\ref{fig:space and time decoding graph}, a more effective parallel system design is essential. Furthermore, circuit-level noise introduces more intricate graph splitting and merging operations across both time and space, which are not fully addressed in previous works~\cite{Tan2022ScalableSD,skoric2023parallel,bombin2023modular}.

\begin{figure}
\begin{center}
    \subfigure[Block Decoding Across Time and Space]{
        \includegraphics[width=\linewidth]{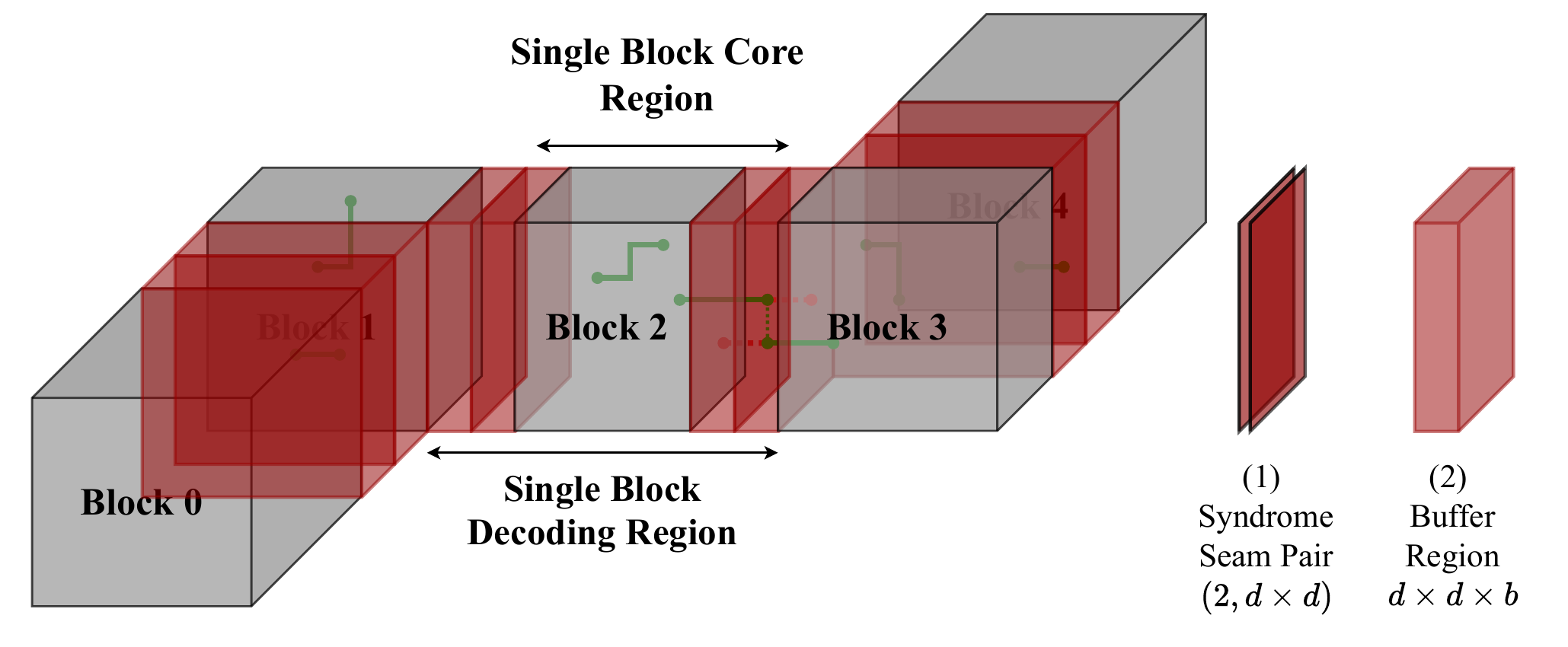}
        \label{fig:block partition}
    }
    \subfigure[Time Block Merging Operation]{
        \includegraphics[width=\linewidth]{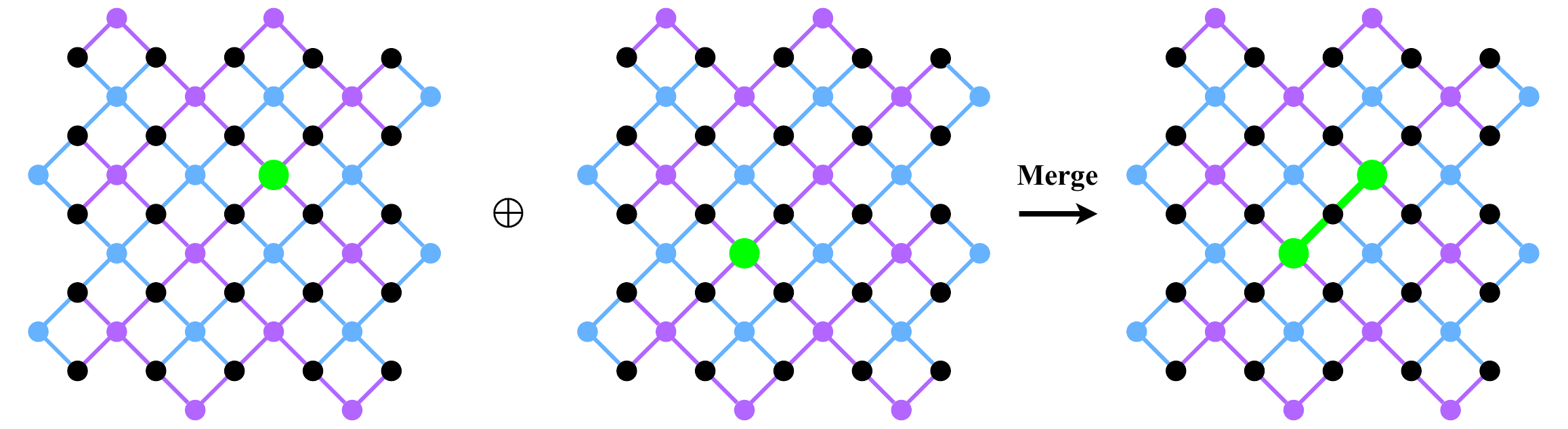}
        \label{fig:time merge}
    }
    \subfigure[Space Block Merging Operation]{
        \includegraphics[width=\linewidth]{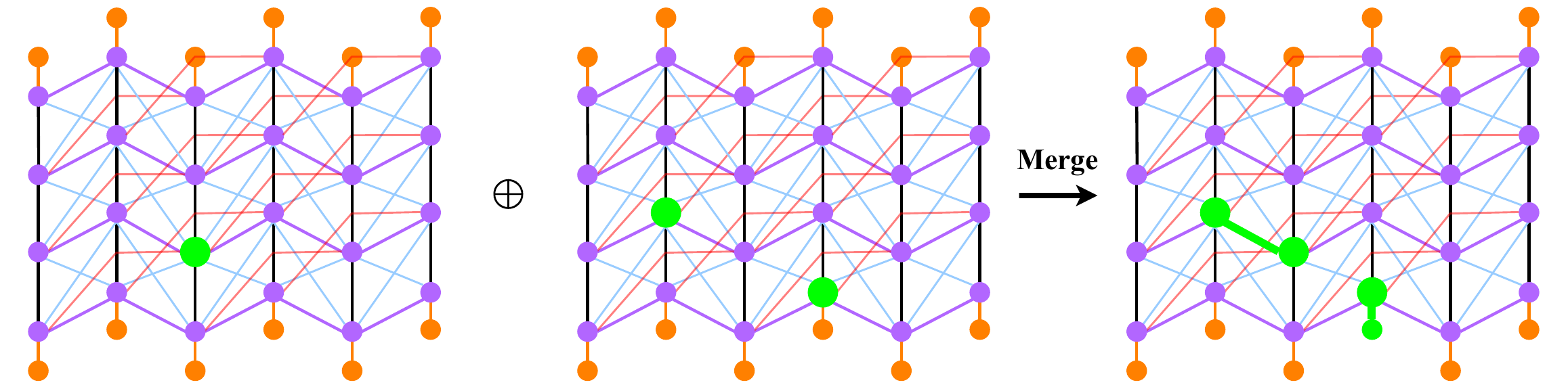}
        \label{fig:space merge}
    }
    \caption{Block decoding and merging. (a) The decoding process decodes a time or space block including the buffer region, while only correcting the core region and leaving corrections at the left/right seams. (b) Time merge operation, with the time axis perpendicular to the plane. (c) Space merge operation. We only plot the Z decoding graph and the time axis is vertically upward. In both (b) and (c), each correction pair of seams between adjacent blocks will go through XOR and 2D decoding to finish the merging process.}
\label{fig:decode and merge}
\end{center}
\end{figure}

To further parallelize existing decoding works and extend them equally across both time and space dimensions, we generalize the time window to the decoding \texttt{block}, which consists of the core region with size $d \times d \times d$ and the buffer region with $b$ time or space layers, as in Figure~\ref{fig:block partition}. We decouple the decoding and merging operations into two independent processes in Algorithm~\ref{alg:Global}, each assigned to the \emph{Decode Thread Pool} (DTP) or the \emph{Merge Thread Pool} (MTP) in Figure~\ref{fig:decoding architecture pipeline}\textcolor{blue}{(a)} for asynchronous execution, with \emph{Global Dynamic Scheduler} in $\S$~\ref{sec:global dynamic scheduler} operating externally. 

Figure~\ref{fig:decode and merge} shows the difference between decoding and merging: The decoding is performed on the \texttt{3D} graph --- the decoding \texttt{block}, while the merging is actually performed \textbf{only} on the \texttt{Zig-Zag} shaped \texttt{2D} graph --- the syndrome seam pair. The decoding buffer and \emph{Merge Mem} allows absolutely asynchronous execution of these two processes, different from interdependent parallelism in~\cite{acharya2025quantum, wu2023fusion}. This provides more concurrency and better performance in decoding latency with near-optimal utilization of computational resources, demonstrating strong scalability.

\subsection{Global Dynamic Scheduler}
\label{sec:global dynamic scheduler}
The overall BLDS in Figure~\ref{fig:decoding architecture pipeline} is managed by a \emph{Global Dynamic Scheduler} described in Algorithm~\ref{alg:Global}. As syndromes are generated every $\mu s$, the scheduler splits them into different blocks virtually through hash-based address mapping, dynamically allocating threads for decoding and merging, and updating the logical states $L_S^G$ for feedback. As long as the syndromes can be processed with sufficient throughput, no memory-overflow will occur.

The streaming decoding pipeline operates as a two-tier producer-consumer model~\cite{dijkstra2002cooperating}, allowing for fully asynchronous execution across different temporal or spatial decoding tasks. Specifically, we employ a more intuitive and robust scheduling method using \textit{P} (wait) and \textit{V} (signal) primitives on semaphores~\cite{dijkstra2002cooperating} and \emph{Merge Mem} to decouple the decode and merge processes. As illustrated in Figure~\ref{fig:decoding architecture pipeline}\textcolor{blue}{(b)}, two producer-consumer mechanisms are employed: (1) The Syndrome Generator acts as \emph{producer}, and the Decode Process acts as \emph{consumer}. (2) The Decode Process acts as \emph{producer} and the Merge Process acts as \emph{consumer}. The syndromes produced by the \emph{Syndrome Generator} are stored into the \emph{Decode Mem}. Once the data reaches the size of one syndrome \texttt{block}, a decode semaphore is released to trigger a decoding thread from the DTP. After a decoding process is completed, the boundary correction — the syndrome seam — is directly stored into the \emph{Merge Mem}. Once a syndrome seam pair is collected, a merge semaphore is then released to trigger a merging thread from the MTP. Therefore, a decoding thread is released back to the DTP after decoding, without waiting for other decoding threads. After all threads finish execution, the \emph{Global Dynamic Scheduler} retrieves the corresponding logical feedback result $L^G_S$ for the next feedback TICK. While we focus on the case of two \texttt{block} neighbors in both Figure~\ref{fig:decoding architecture pipeline}\textcolor{blue}{(b)} and Algorithm~\ref{alg:Global} for clarity, the approach remains consistent for one or three neighbors in more complex cases~\cite{litinski2019game,tan2024sat}. All scheduling across different decoding or merging threads is managed by the scheduler, thus finishing the FTQC program in a fully streaming pipeline.

\begin{algorithm}
\caption{Global Scheduling Strategy}
\label{alg:Global}
    \KwIn{Sparse syndrome data $S^\prime$ from the AXI BUS}
    \KwOut{Updated global logical states $L^G_S$ at feedback TICK}
    \textbf{Initialize:} Global logical states $L^G_S$; Logical boundary set $L_B$; Decode and Merge Mem $M_D, M_M$; Decode and Merge Queue $Q_D, Q_M$; Decode and Merge Thread Pool $DT[M],MT[N]$; $i = 0$\

    \While {i $<$ Total\_Block\_NUM} {
        \textbf{Run} $DT[M],MT[N]$\;
        \textbf{Collect} Syndrome data after local decoding $S^\prime$\;
        \textbf{Partition} $S^\prime$ : $S^\prime \to \{B_i\}$\; 
        \textbf{Map} Syndromes to Decode Mem: $B_{i} \to M_D[Hash(i)]$\;
        $Q_D.push(i)$, Semaphore[0].V()\;
        % \For{$j$ in Neighbor($B_i$)}{
        %     \If{(Finished ($B_i$ and $B_{j}$))}{
        %         $Q_M.push((i, j))$, Semaphore[1].V()\;
        %     }
        % }
        \If{Feedback TICK}{
            \textbf{Send} $L^G_S$ $\to$ Control Hardware\;
        }
        $i = i + 1$\;
    }
    \SetKwProg{Procedure}{Procedure}{}{}
    \SetKwProg{myproc}{Procedure}{}{}
    \myproc{\textnormal{\Call{Decode}{thread\_id}}}{ 
    Semaphore[0].P() \;
    $ i \gets Q_D.front(), Q_D.pop()$\;
    $\mathcal{E}$ = Decode($M_D[Hash(i)]$)\;
    \If{$\mathcal{E} \cap L_B$}{
        Update $L^G_S$\;
    }
\textbf{Save} $\mathcal{E} \to M_M[Hash(i)][0],M_M[Hash(i)][1]$\;
    \For{$j$ in Neighbor[$i$]}{
        \If{(Finished[$i$] \& Finished[$j$])}{
            $Q_M.push((i, j))$, Semaphore[1].V()\;
        }
    }
}
\myproc{\textnormal{\Call{Merge}{thread\_id}}}{
    Semaphore[1].P() \;
    $(i,j) \gets Q_M.front(), Q_M.pop()$\;
    $S^0,S^1 = M_M[Hash(i)][1],M_M[Hash(j)][0]$\;
    $\mathcal{E}$ = Decode($S^0\oplus S^1$)\; 
    \If{$\mathcal{E} \cap L_B$}{
        Update $L^G_S$\;
    }
}
\end{algorithm}

\section{Neural Local Decoding Unit}
\label{sec:neural local decoding unit}

Superconducting qubits on chip are read out via microwave signals processed by FPGA-based digitizers~\cite{walter2017rapid, yang2022fpga, xu2021qubic}. This enables the distributed integration of local decoding units. Accordingly, we design the neural local decoding unit (NLDU) for seamless integration with readout hardware on FPGAs, allowing preprocessing and error correction before syndrome uploading.

In the following subsections, we will describe our key ideas to design a super-light local decoding unit with the ability to implement on the FPGA and execute in a streaming manner.

\subsection{Syndrome Embedding and Error Mapping}
\label{sec:Error Decomposition And Mapping}
\begin{figure}[ht]
\begin{center}
    \subfigure[Syndrome embedding]{
        \includegraphics[width=\linewidth]{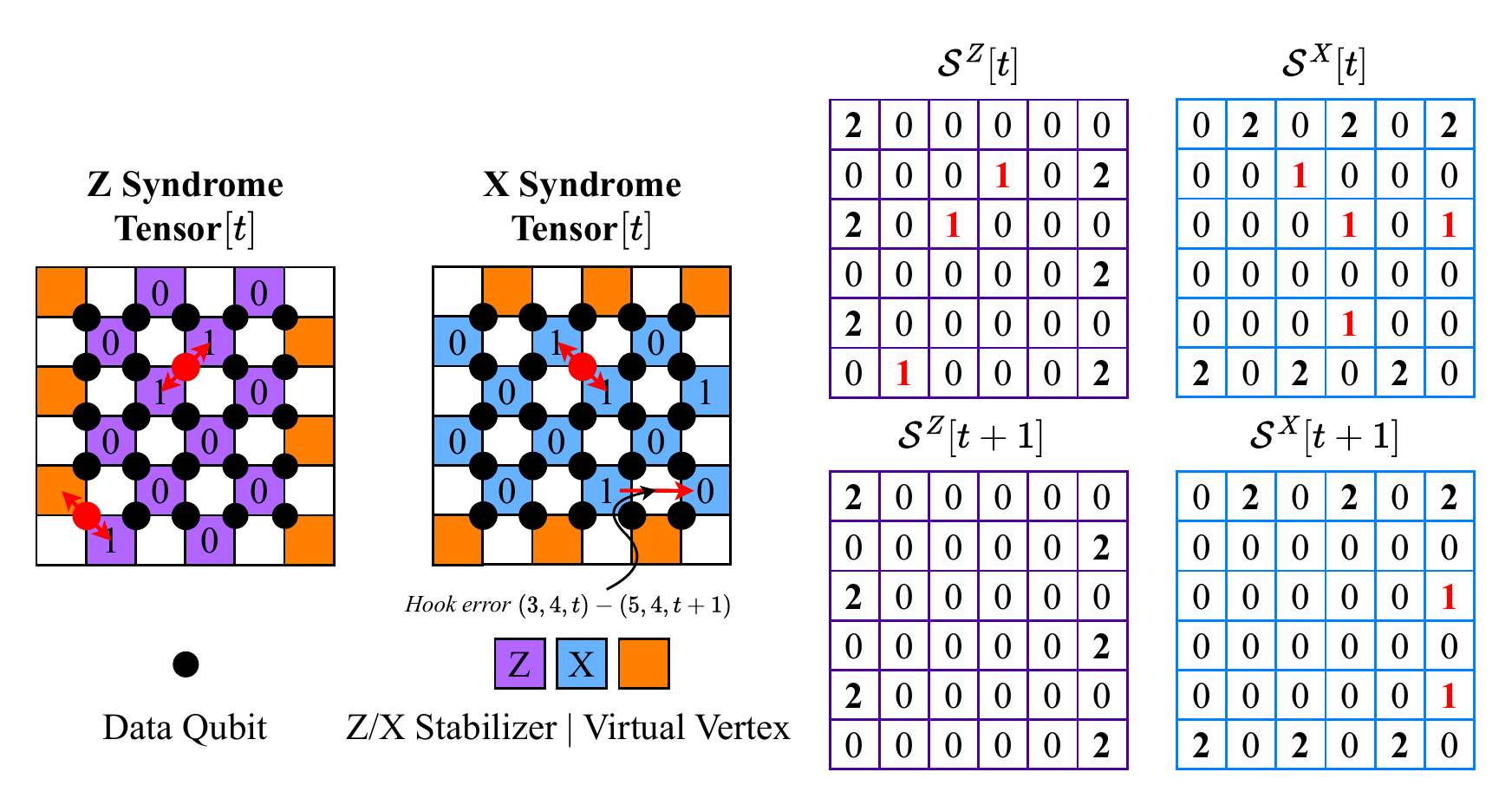}
        \label{fig:NLDU input}
    }
    \subfigure[Error decomposition and mapping]{
        \includegraphics[width=\linewidth]{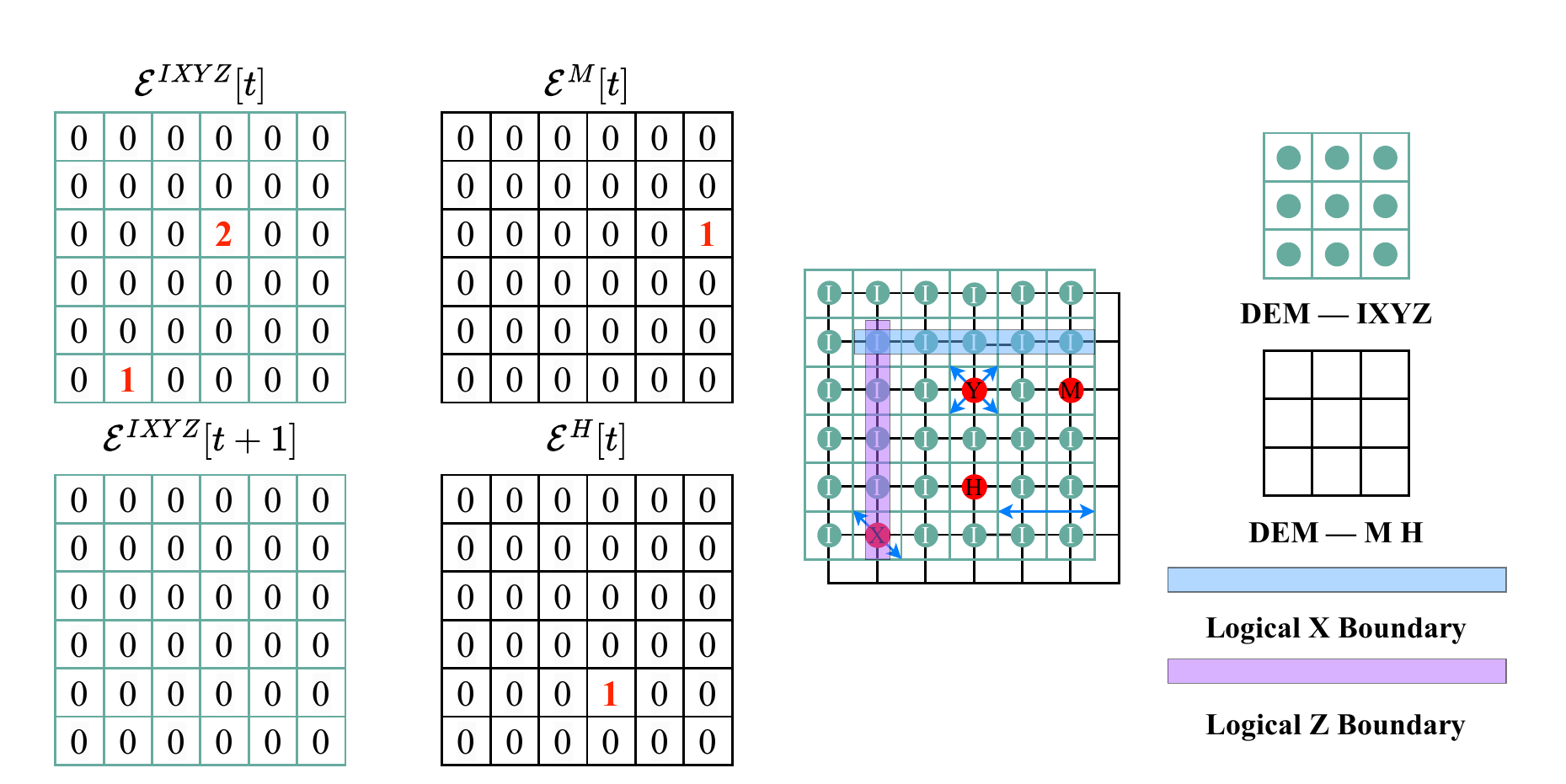}
        \label{fig:NLDU output}
    }
    \caption{Tensor representation for $d = 5$ surface code. (a) The syndrome information is embedded into the $\mathcal{S}^Z, \mathcal{S}^X$ tensors. (b) The ground truth error labels before one-hot encoding are shown in $\mathcal{E}^{IXYZ}, \mathcal{E}^{M}$ and $\mathcal{E}^{H}$.}
    % \label{fig:nn decoder input and output}
\end{center}
\end{figure}

As described in $\S$~\ref{sec:surface_code}, syndrome outcomes of adjacent measurement rounds form detectors. To represent the X and Z syndrome detectors $\{\mathcal{S}^X, \mathcal{S}^Z\}$ simultaneously, we adopt the coordinate system from Stim~\cite{Gidney2021StimAF} to create a novel tensor representation. For instance, in a $d \times d$ rotated surface code patch illustrated in Figure~\ref{fig:NLDU input}, the coordinates of all X and Z detectors, as well as boundary virtual vertices, fit into a $(d+1, d+1)$ matrix. We further extend this by placing $\mathcal{S}^X$ and $\mathcal{S}^Z$ in separate channels to distinguish them, i.e. $\mathcal{S} = [\mathcal{S}^X \ \mathcal{S}^Z]$. Consequently, the syndrome information $S$ for arbitrary $\alpha \times \beta$ logical patch with $\gamma$ measurement rounds can be  represented as a tensor with shape $(\alpha+1, \beta+1, \gamma, 2)$. Here a value of $1$ indicates a defect, while a value of $0$ represents either a vertex without a defect or a coordinate that does not correspond to a vertex of the specified type in the decoding graph. To distinguish logical boundaries, the syndrome tensor also includes a constant value of $2$, indicating virtual vertices on the boundary~\cite{Tan2022ScalableSD}.

The output of the neural network represents the predicted probabilities of errors in the circuit, i.e. edges in the detector error model (\emph{DEM})~\cite{Dennis_2002, Gidney2021StimAF}. Representing errors becomes more complicated, especially in the circuit-level noise model due to the variety of possible noise types. However, we can still map each error on the $(\alpha, \beta)$ lattice with $\gamma$ rounds to a unique position in a $(\alpha+1, \beta+1, \gamma)$ space, associated with $6$ distinct error channels, by carefully analyzing and processing the errors in the \emph{DEM}: 
\begin{itemize}
    \item {\emph{Horizontal Edge}}: Each horizontal edge has a well-defined spatial location, namely the location of the corresponding data qubit. For every round of detectors, on each data qubit, there can be one Pauli error from $\{I, X, Y, Z\}$, where $I$ means no error. Therefore we treat this as a classification problem with $4$ classes, and use $4$ channels labeled $\{I, X, Y, Z\}$ to represent horizontal edges.
    \item {\emph{Vertical Edge}}: Each vertical edge is easily mapped to the location of the temporally earlier end point of that edge. We use a separate channel labeled $\{M\}$ to represent vertical edges, i.e. measurement errors.
    \item {\emph{Diagonal Edge}}: Each diagonal edge can be \emph{decomposed} into a vertical edge and a horizontal edge, i.e. $M + X \ or \ M + Z$ to avoid ``right'' or ``left'' confusion. Although such a decomposition may create equivalent errors~\cite{Dennis_2002}, as endpoints of both edges are in the receptive field, the NN should be able to take this correlation into account and decode correctly.
    \item {\emph{Hook Edge}}: Each hook edge is handled similar to vertical edges, i.e. mapped to the location of the first end point of that edge, with a separate channel labeled $\{H\}$ , representing a hook error caused by quantum error propagation in stabilizer measurement process.
\end{itemize}

For each edge type, there will be some locations near the logical boundary that are not mapped to any edge of that type, predictions at those locations are ignored. After error decomposition and mapping, the output channel size should be $6$, i.e. $\mathcal{E} = [\mathcal{E}^{I} \ \mathcal{E}^{X} \ \mathcal{E}^{Y} \ \mathcal{E}^{Z} \ \mathcal{E}^{M} \ \mathcal{E}^{H}]$, corresponding to $3$ classification problems. Thus, through a tensor $\mathcal{E}$ in the shape of $(\alpha+1,\beta+1,\gamma,6)$, we perfectly represent all kinds of errors in the \emph{DEM} as Figure~\ref{fig:NLDU output}.

% \begin{equation}
%     \mathcal{E} = \begin{bmatrix} \mathcal{E}^{I} & \mathcal{E}^{X} & \mathcal{E}^{Y} & \mathcal{E}^{Z} & \mathcal{E}^{M} & \mathcal{E}^{H} \end{bmatrix}
%     \label{formula:DEM representation}
% \end{equation}

With this novel tensor representation, we have transformed syndrome and error information of arbitrary lattice~\cite{litinski2019game} into a tensor with a consistent topological structure, preserving all relevant details in the decoding hypergraph, which is discussed in $\S$~\ref{sec:quantum error pattern} before.

\subsection{Optimization Goal}
Now the local decoding problem is transformed from a subgraph matching problem into a vertex-by-vertex DEM multi-classification problem, with the loss function below:

% \begin{equation}
\[
    % \begin{align}
    % \mathcal{L}(\Theta) =\ 
    % & - \sum_{i=1}^{4} e^{P_i} \log(\hat{e}^{P_i}) \notag \\
    % & - \left[ e^M \log(\hat{e}^M) + (1 - e^M) \log(1 - \hat{e}^M) \right] \notag \\
    % & - \left[ e^H \log(\hat{e}^H) + (1 - e^H) \log(1 - \hat{e}^H) \right]
    % \end{align}
    \begin{aligned}
        \mathcal{L}(\Theta(\mathcal{S})) = &-\sum_{P \in \{I, X, Y, Z\}} e^{P} \log(\hat{e}^{P}) \\
        & - \sum_{V \in \{M, H\}} \left[ e^V \log(\hat{e}^V) + (1 - e^V) \log(1 - \hat{e}^V) \right]
        % - \sum_{i=0}^{1} (e^{M} \log(\hat{e}^{M}) + e^{H} \log(\hat{e}^{H}))
    \end{aligned}
\]
% \label{formula:optimization goal}
% \end{equation}
where $\{e^I, e^X, e^Y, e^Z, e^M, e^H\}$ are the ground truth labels from the \emph{DEM}, $\{\hat{e}^I, \hat{e}^X, \hat{e}^Y, \hat{e}^Z, \hat{e}^M, \hat{e}^H\}$ are the predicted error probabilities. The neural network parameterized by $\Theta$ maps $\mathcal{S}$ to $\hat{\mathcal{E}}, \hat{e} \in \hat{\mathcal{E}}$.

We address this problem using a fully convolutional neural network~(FCNN) in Table~\ref{tab:model arch}, capable of accepting inputs of any size and predicting in parallel. The model is quite light-weight, with approximately 3,000 parameters only. In particular, its receptive field ($R \times R \times R = 7 \times 7 \times 7$) covers up to 1 hook error or 2 data qubits errors, aligning with the motivation from $\S$~\ref{sec:quantum error pattern}.
% \begin{scriptsize}
\begin{table}[h!]
    \centering
    \small
    \caption{FCNN Architecture of The NLDU}
    \label{tab:model arch}
    \begin{tabular}{ccccc}
        \toprule
        \textbf{Layer} & \textbf{Network} & \textbf{\#In-Ch.} & \textbf{\#Out-Ch.} & \textbf{Kernel Size} \\
        \midrule
        1 & Conv3d + \emph{ReLU} & 2 & 7 & $3\times 3 \times 3$ \\
        2 & Conv3d + \emph{ReLU} & 7 & 7 & $3\times 3 \times 3$ \\
        3 & Conv3d + \emph{ReLU} & 7 & 7 & $3\times 3 \times 3$  \\
        4  & Conv3d & 7 & 6 & $1\times 1\times 1$ \\
        \bottomrule
    \end{tabular}
\end{table}
% \end{scriptsize}

\subsection{Post-Processing and Hierarchical Workflow}
Figure~\ref{fig:NLDU input}~\ref{fig:NLDU output}  shows an example of the prediction and post-processing of the NLDU: Each \emph{DEM} error ``position'' has 6 channels, representing the $\{X, Y, Z\}$ errors occurring on the data qubit at the corresponding position or $\{M, H\}$ errors mapping to the right down stabilizer qubit. 
Suppose there are 4 errors (red arrows) occurring during rounds $t$ and $t + 1$, there should be 4 error predictions (blue arrows): $\{X, Y, M, H\}$. For $\{X, Y\}$ errors, i.e. $\{(0,4,t)-(1,5,t),(2,1,t)-(3,2,t),(2,2,t)-(3,1,t)\}$, we remove the syndromes from $S$ and record a a flip of the logical $Z$ in the $L^L_S$ since the $X$ error occurs on the logical $Z$ boundary. For $\{M, H\}$ errors, the outputs represent a measurement and hook error between rounds $t$ and $t+1$, we correct them similarly. In the end, the remaining syndrome data $S^\prime$ needs to be uploaded to the BLDS via AXI bus, while the local logical states $L^L_S$ are reserved until the feedback TICK, together with $L^G_S$ from the BLDS to determine the next logical operation.

\section{System Implementation on FPGA-CPU Hybrid Architecture}
\label{sec:system implementation}

\subsection{CPU Implementation}
The implementation of BLDS on CPU is based on a dual-socket \emph{Intel Xeon Gold 6248R} CPU, featuring 24 cores per socket, 2 threads per core, a base frequency of 3.00 GHz, and a maximum frequency of 4.00 GHz. We implemented the BLDS using \emph{C++} and utilize \emph{Sparse Blossom}~\cite{higgott2025sparse} as the base decoder. Due to the lack of existing work for automatically generating decoding graphs in lattice surgery, we manually constructed various temporal or spatial decoding graphs using Pymatching~\cite{higgott2022pymatching} and Stim~\cite{Gidney2021StimAF} for experimental benchmarking. More automated approaches are worth exploring in the future.

\begin{figure}[htbp]
\centering
\includegraphics[width=\linewidth]{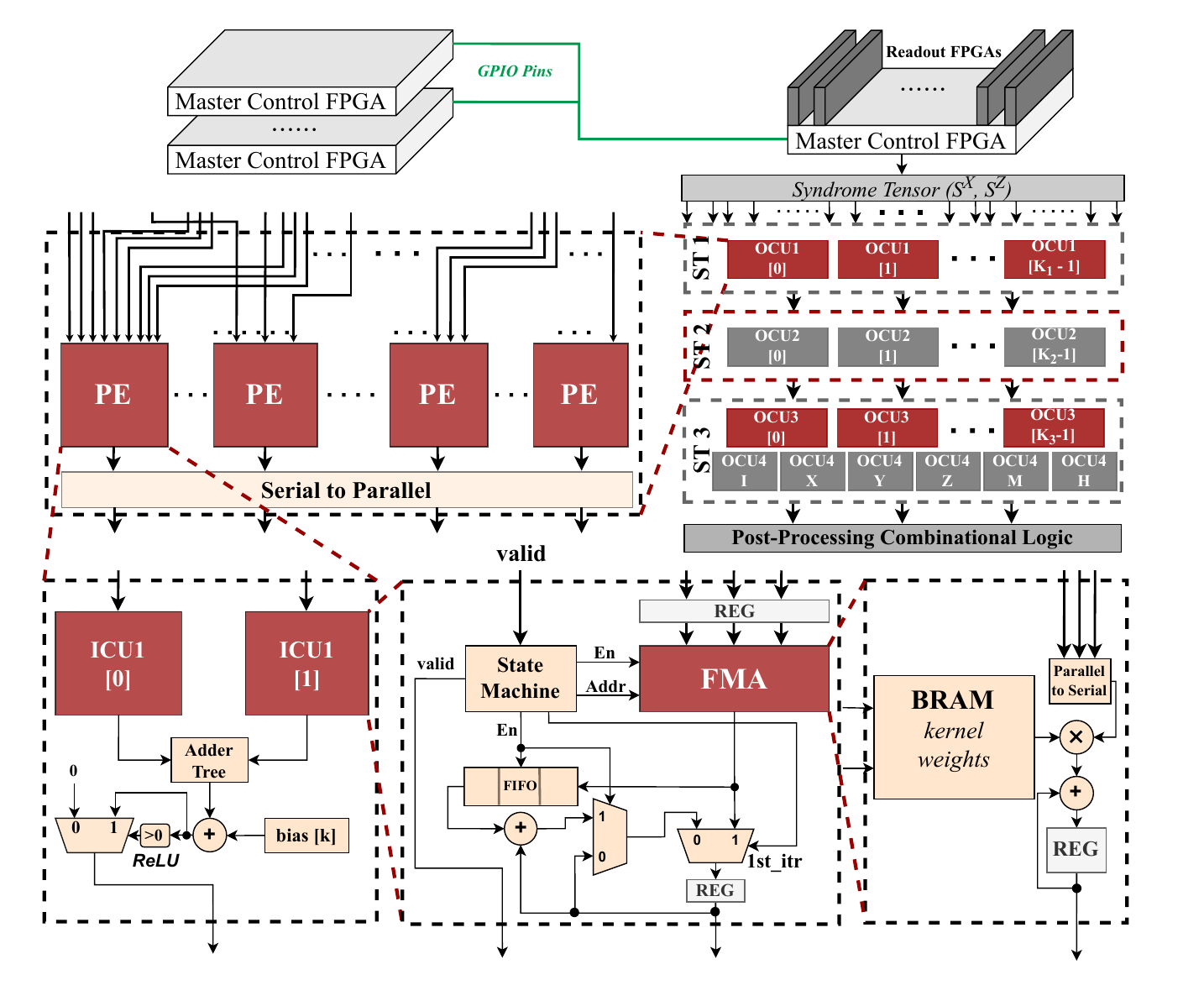}
\caption{Microarchitecture of the NLDU. The input of each layer at time step $t$ and each convolution kernel are represented as $In_{i, j, t, l}$ and $kernel_{x, y, z, l, k}$, where $l$ and $k$ denote the input and output channels. Each \texttt{OCU[k]} computes the convolution for output channel $k$ and each \texttt{ICU[l]} computes $\sum_{x=0}^2\sum_{y=0}^2\sum_{z=0}^2In_{i+x,j+y,t+z,l}\cdot kernel_{x,y,z,l,k}$.}
\label{fig:NLDU architecture}
\end{figure}

\subsection{FPGA Implementation}
\label{sec:FPGA_implementation}
For the NLDU, we first trained the quantization-aware INT8 model on CPU, and then synthesized and implemented the hardware architecture on \emph{Xilinx Alveo U200}~\cite{amd2024} boards, each of which has 35MB on-chip memory (1,766 36Kb BRAM, 960 288Kb URAM), 1,182K LUTs, 2,364K registers and 6,840 DSP slices. 

\subsubsection{Model Quantization}
The lightweight of the model and ternary characteristics of input values enable the network in NLDU to be easily quantized. We adopt quantization-aware training~\cite{jacob2018quantization, NEURIPS2019_9015} which models the effects of quantization during training, allowing for high accuracy close to the full-precision model. We use $d = 13$ circuit-level noise data generated by Stim~\cite{Gidney2021StimAF} to train the INT8 model, as detailed in Table~\ref{tab:model training}. The training approach enables accurate local decoding across different $d$ and $p$, due to the translation invariance of convolutional networks. After approximately only 10~\textit{min} of training from scratch, the NLDU achieves high accuracy across all $6$ error types. The false positive (FP) and false negative (FN) rates reach $0.0189\%$ and $6.85\%$ for Pauli errors, $0.0059\%$ and $7.18\%$ for measurement errors, and $0.0008\%$ and $16.18\%$ for hook errors, respectively. The FN rates are higher due to the local decoding strategy: predictions with low confidence are transmitted back to the BLDS for further decoding. These results correspond to overall classification accuracies above $99.9\%$, demonstrating the model’s rapid convergence and robust error detection capability.

\begin{table}[htbp]
    \centering
    \small
    \caption{Quantization-Aware Training on CPU}
    \label{tab:model training}
    \begin{tabular}{cccc}
        \toprule
        \textbf{Training Step} & \textbf{I} & \textbf{II} & \textbf{III} \\
        \midrule
        $p$ & 0.001 & 0.003 & 0.005 \\
        Batch Size $\times$ Epoch & 32 $\times$ 20000 & 32 $\times$ 20000 & 32 $\times$ 20000 \\
    \bottomrule
    \end{tabular}
\end{table}
% We observed almost same accuracy as the FP32 model.
\subsubsection{3-Stage Streaming Inference}

% Streaming Pipeline FMA and Adder Tree, FIFO, 
Since the syndromes are read out continuously at $1\mu s$ per round, and we also want the NLDU-to-BLDS output to follow the same feature, a fully streaming inference pipeline is proposed to process the syndromes in real time, which differs from classical CNN inference frameworks and represents the first streaming local decoding architecture that operates seamlessly without disrupting the original syndrome transmission pattern. The NLDU is divided into 3 stages corresponding to Table~\ref{tab:model arch}: \textbf{ST1}---\{layer1\}, \textbf{ST2}---\{layer2\} and \textbf{ST3}---\{layer3 + layer4\}, 3rd and 4th layers are combined since the last layer only performs a localized operation. To enable the streaming inference without backlog, it's important to make sure the latency of each stage is under $1\mu s$. 

The microarchitecture of the NLDU is depicted in Figure~\ref{fig:NLDU architecture}, with \textbf{ST1} shown in detail without loss of generality. Convolutions corresponding to distinct output channels are mapped to individual \texttt{OCU} blocks, each comprising multiple \texttt{PE}s that execute parallel computations across input channels.
To enable streaming inference, each \texttt{ICU} in \texttt{PE} collects data from a distinct input channel from 3 consecutive $\mu s$ to compute a single valid output. Each data at time step $t$ is multiplied 3 times by different weights using the \texttt{FMA} and then stored into the \texttt{FIFO}. At each $\mu s$, $\sum_{x=0}^2\sum_{y=0}^2 In_{i+x, j+y, t, l}\cdot kernel_{x, y, 2, l, k}$ is first calculated by the \texttt{FMA} and stored in the \texttt{REG}. The $In$ is then multiplied by $kernel_{x, y, 1, l, k}$ and $kernel_{x, y, 0, l, k}$ in order, where the results are stored into the \texttt{FIFO}. Meanwhile, the leftmost elements in the \texttt{FIFO} are popped out and accumulated with the \texttt{REG} to get the valid output. There is also a \texttt{State Machine} to generate the control and enable signals that manage the memory access, govern the data flow, and handle the boundary conditions. All parameters of the NN are stored in the \texttt{BRAM}.

At each $\mu s$, \textbf{ST3} outputs a tensor of shape $(N, N, 6)$, where $N \times N$ denotes the patch size managed by a single FPGA. Each vertex has $6$ outputs: 1-4 represent the probability of occurring a Pauli error, i.e. $\{I, X, Y, Z\}$, 5 or 6 represents the probability of occurring a measurement error $\{M\}$ or a hook error $\{H\}$, respectively.

\begin{figure}[htbp]
\centering
\includegraphics[width=\linewidth]{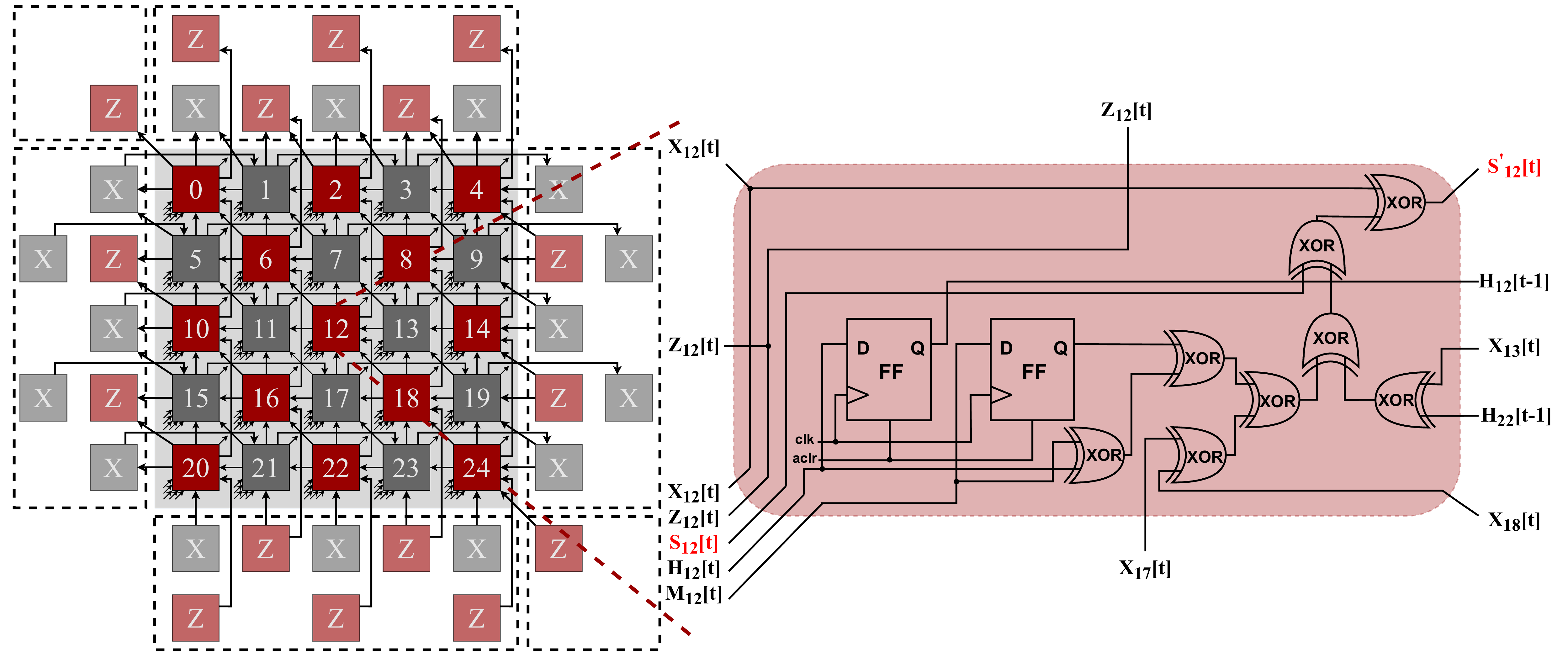}
\caption{Combinational logic for syndrome updating of the NLDU. For visulization, we illustrate the case where each master FPGA controls a $5\times 5$ qubit patch. The red and grey blocks represent the Z and X syndrome vertices. Only the combinational circuit for the Z syndrome vertex is shown here; the circuit for the X vertex is similar, with a different hook error direction. To support distributed deployment for larger patches, error predictions from neighboring FPGA boards (semi-transparent regions) are also required.}
\label{fig:parallel post-processing}
\end{figure}

\subsubsection{Parallel Post-Processing Combinational Logic}
The 3-stage inference output can be post-processed in parallel using a carefully designed combinational logic: 

\textbf{Virtual Dequantization.} There is no need of real dequantization for \textbf{ST3}'s output: (1) For \textbf{ST3}[:, :, 0:4], we only need to choose the class with the maximum probability using a \texttt{Comparator Tree}. (2) For \textbf{ST3}[:, :, 4] and \textbf{ST3}[:, :, 5], which represent two binary classification probabilities, there is no need of \emph{Sigmoid} to get the confidence $C_{M}$ and $C_{H}$. Instead, we set the FP32 threshold $\theta = 0.8$ and calculated the corresponding pre-\emph{Sigmoid} threshold: $1 / (1 + e^{-\tilde{\theta}}) = 0.8, \tilde{\theta} = ln(4)$. The $\tilde{\theta}$ is then transformed into the fixed-point threshold $\theta_{INT8}$, once \textbf{ST3}[:, :, 4] or \textbf{ST3}[:, :, 5] $> \theta_{INT8}$, we consider that one measurement error or hook error has occurred.
 
\textbf{Parallel Syndrome Update.} To facilitate the parallel post-processing, we compress the output results into 4 channels, $E = [X,Z,M,H]$. If one DEM position in Figure~\ref{fig:NLDU output} has one $Y$ error and one hook error, $E$ should be $[1,1,0,1]$. After that, the syndrome and logical updating can be paralleled in Figure~\ref{fig:parallel post-processing}. For example, $S^\prime_{12}[t]$ after NLDU is affected only by its neighboring predictions:
% \begin{equation}
\[
    \begin{split}
    S^\prime_{12}[t] &= S_{12}[t] \oplus M_{12}[t] \oplus M_{12}[t-1] \oplus H_{12}[t] \\
    &\oplus H_{22}[t-1] \oplus X_{12}[t] \oplus X_{13}[t] \oplus X_{17}[t] \oplus X_{18}[t]
    \end{split}
    \label{formula: syndrome outcome}
\]
% \end{equation}
as the syndrome outcome at each stabilizer will be affected only by the data qubits connected, or the related measurement and hook errors. The same applies to the updating of the logical states $L_S^L$.

\subsubsection{Cross-Board and FPGA-CPU Communication}
For near-term quantum chips~\cite{acharya2025quantum, gao2025establishing}, one FPGA board is sufficient for the demonstration of logical quantum memory experiment. While for large-scale quantum computing scenarios, our NLDU implementation is modular and suitable for a distributed deployment, which requires: 

(a) Pre-Inference Broadcasting: Although global syndrome data is not required for the FCNN, neighboring measurement data ($(3+N+3)^2 - N^2$) from other FPGA boards is still needed due to the receptive field as $R = 7, \lfloor R / 2\rfloor = 3$. Considering the fixed transmission pattern and low information density, we directly broadcast neighboring syndrome data between FPGA boards via GPIO pins before the \textbf{3-stage Inference}, as shown in Figure~\ref{fig:NLDU architecture}.

(b) Post-Inference Updating: The distributed NLDU also requires neighboring $[X,Z,M,H]$ to finish the syndrome updating combinational logic in Figure~\ref{fig:parallel post-processing}. The communication interaction can also be accomplished using GPIO pins, with a state machine to achieve sub-$\mu s$ level data synchronization. All syndrome uploading and logical feedback of $L^G_S$ is implemented via the AXI4-Lite-Slave interface.

\subsubsection{Configurable and Flexible IP Generator}
\label{sec:configurebale IP generator}
The architecture of the NLDU in Figure~\ref{fig:NLDU architecture} is fully blockized. To avoid throughput bottlenecks, it is sufficient to ensure that the latency of each inference stage remains below $1\mu s$. Consequently, as long as this constraint is met, certain hardware blocks can be reused to reduce resource consumption, via a state machine that manages circular block sharing and effectively converts parallelism into serialized reuse. In our design, the patch size $N \times N$ readout processed by a single FPGA, the number of \texttt{OCU} --- $K_i$ and the number of parallel \texttt{PE} units --- $P_i$ are both configurable ($K_0 = 2$ is fixed). We derive empirical formulas in Equations~\eqref{formula:all_estimations} to model the usage of LUTs, and registers (REG), as well as the NLDU latency (LTC) from the syndrome input to its prediction output during stream processing. These estimations are based on $N$, ${K_i}$, $P_i$, and the clock frequency $f$:

\begin{subequations} \label{formula:all_estimations}
\begin{align}
    % \text{DSP} &\approx \sum\nolimits_{i=1}^37P_iK_{i-1}+6K_3N^2;  \label{formula:Hardware estimation DSP}\\
    \text{LUT} &\approx \sum\nolimits_{i=1}^37P_i(40K_{i-1}+1)+16K_3N^2; \label{formula:Hardware estimation LUT}\\
    \text{REG} &\approx \sum\nolimits_{i=1}^3 56P_i(1+K_{i-1})+16K_3N^2; \label{formula:Hardware estimation REG}\\
    \text{LTC} &\approx 3\mu s+\frac{28}{f}\bigg(3+\sum\nolimits_{i=1}^3(\lceil(N+2(3-i))^2/P_i\rceil)\bigg). \label{formula:Hardware estimation LATENCY}
\end{align}
\end{subequations}
where the $3\mu s$ latency is due to the fixed pipeline delay during the 3-stage inference. We also develop an IP generator to explore a small design space per layer defined by $K_i$ and $P_i$, under latency $< 1\mu s$ per stage. We evaluate up to 90 configurations each layer and select the one with the lowest resource cost. RTL is then generated using \emph{Python} + \emph{Verilog} for regular modules and \emph{SpinalHDL} for complex ones. As the NLDU is localized with constant $N$, the design space remains compact regardless of code distance.

\section{Evaluation}
\label{sec:evaluation}

We first present a qualitative comparison comparing LATTE with existing hardware-implemented decoding architecture in Table~\ref{tab:decoder_comparison} and then systematically benchmark the key metrics --- \emph{accuracy}, \emph{transmission bandwidth}, \emph{throughput}, and \emph{latency} --- of LATTE using different and sufficient experiments. To ensure the fairness of experimental evaluations, we used Stim~\cite{Gidney2021StimAF} to simulate all the circuits under the the same circuit-level noise model as in~\cite{Tan2022ScalableSD} across different experiments. Finally, we report the hardware utilization of our system.

\begin{table*}[ht]
\centering
\caption{Comparison and Tradeoffs of Hardware-Implemented Decoding Architectures}
% \begin{tabular}{|p{1cm}|p{1cm}|p{1cm}|p{1cm}|p{1cm}|p{1cm}|p{1cm}|p{1cm}|p{1cm}|}
\begin{tabular}{|c|c|c|c|c|c|c|c|}
\hline
\textbf{Architecture} & \textbf{Predecode} & \textbf{Streaming$^1$} & \textbf{Accuracy} & \textbf{Bandwidth Reduction} & \textbf{Throughput$^1$} & \textbf{Latency} & \textbf{Scalability}   \\
\hline
\hline
LATTE                                           & \textbf{Yes}      & \textbf{Yes} & Sufficient         & \textbf{High}      & \textbf{High}   & \textbf{Low}        &\textbf{High} \\
Google~\cite{acharya2025quantum, wu2023fusion}  & No                & \textbf{Yes} & $\sim$ MWPM        & Low                & \textbf{High}   & Medium              & Medium       \\
LILLIPUT~\cite{das2022lilliput}                 & No                & \textbf{Yes}           & $\sim$ MWPM        & Low                & \textbf{High}             & \textbf{Low$^3$}    & Low$^4$      \\
AFS~\cite{das2022afs}                           & No                & No           & $\sim$ UF          & Low                & ---             & \textbf{Low$^3$}    & Medium$^4$   \\
CC~\cite{barber2025real}                        & No                & No           & $\sim$ UF          & Low                & ---             & \textbf{Low$^3$}    & Medium$^4$   \\
Promatch~\cite{alavisamani2024promatch}         & \textbf{Yes}      & No           & Medium             & Medium             & ---             & \textbf{Low$^3$}    & Low$^4$      \\
AstreaG~\cite{Vittal2023AstreaAQ}               & \textbf{Yes}      & No           & Low                & Medium             & ---             & \textbf{Low$^3$}    & Low$^4$      \\
CLIQUE~\cite{ravi2023better}                    & \textbf{Yes}      & No           & $\sim$ MWPM$^2$    & Low                & ---             & \textbf{Low$^3$}    & Low$^4$      \\
Lazy~\cite{delfosse2020hierarchical}            & \textbf{Yes}      & No           & $\sim$ MWPM$^2$    & Low                & ---             & \textbf{Low$^3$}    & Low$^4$      \\   
\hline
\end{tabular}
\begin{flushleft}
\footnotesize
$^1$~Streaming means the architecture supports a streaming decoding mode, and throughput measures the ability to decode arbitrary syndrome measurement rounds without backlog. 

$^2$~They degrade into a MWPM decoder at the $p\sim10^{-3}$ circuit-level noise regime, or for large code distance and more syndrome measurement rounds.

$^3$~Low latency for $d \times d \times d$ decoding scenario only.

$^4$~They all target a fixed $O(d^3)$ memory experiment decoding scenario. ~\cite{das2022lilliput} can only deal with $d = 3, 4, 5$, ~\cite{das2022afs, barber2025real} are scalable to more code distances, ~\cite{alavisamani2024promatch, Vittal2023AstreaAQ} operate well only at extremely low error and small code distances regime $p\sim10^{-4}$ and $d = 11, 13$, ~\cite{ravi2023better, delfosse2020hierarchical} are no longer useful at $p\sim10^{-3}$, large $d$ and syndrome measurement rounds $\geq 2$.

\end{flushleft}
\label{tab:decoder_comparison}
\end{table*}

\subsection{Accuracy}
The accuracy of our decoding architecture is evaluated on single-block and multi-block decoding tasks.

\subsubsection{Memory and Stability Experiment}
\begin{figure}[htbp]
\begin{center}
\subfigure[Memory and Stability Experiment]{
\includegraphics[width=\linewidth]{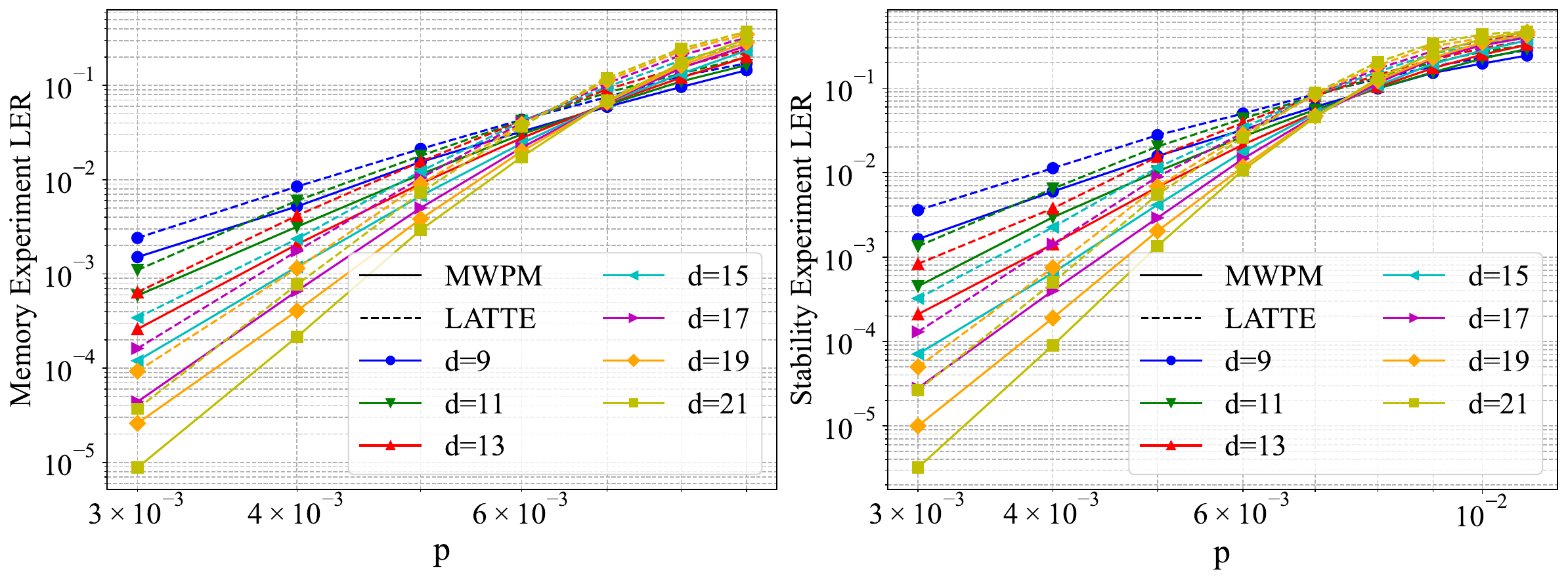}
\label{fig:memory experiment and stability experiment}
}
\subfigure[Temporal and Spatial Multi-Block Decoding Accuracy]{
\includegraphics[width=\linewidth]{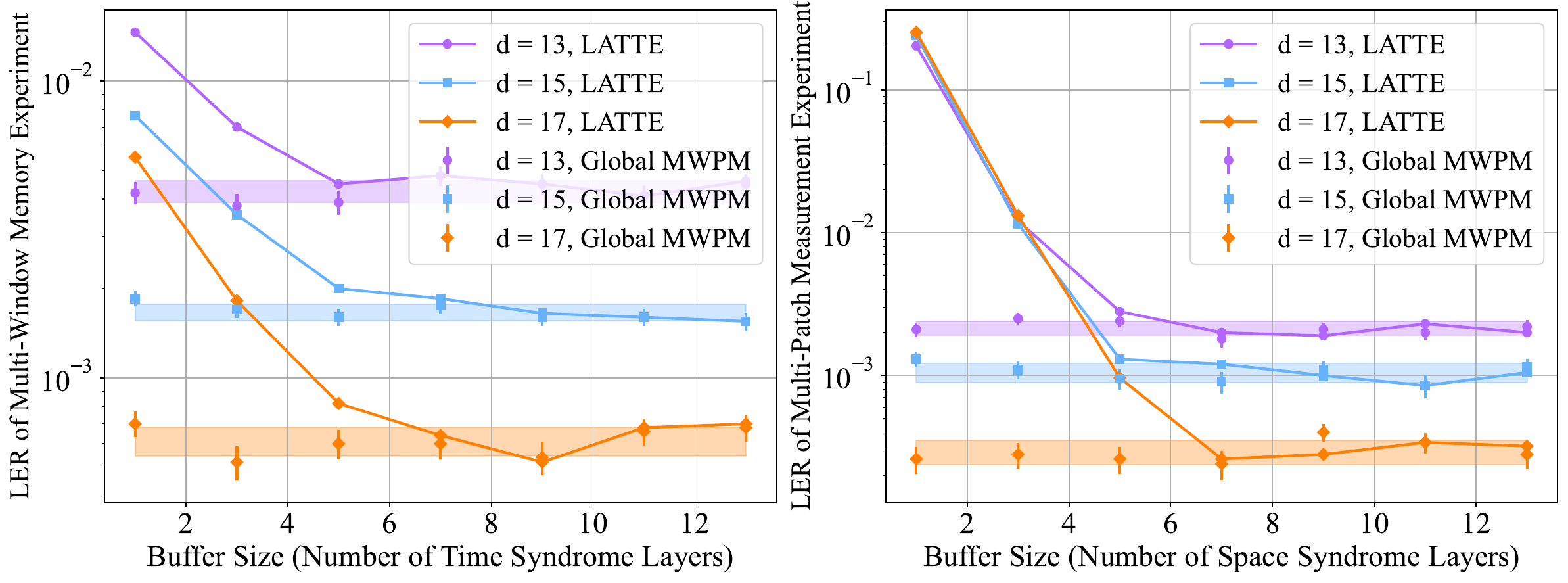}
\label{fig:time and space buffer size experiment}
}
\subfigure[Comparison of Sensitivity in Logical Error Rate]{
\includegraphics[width=\linewidth]{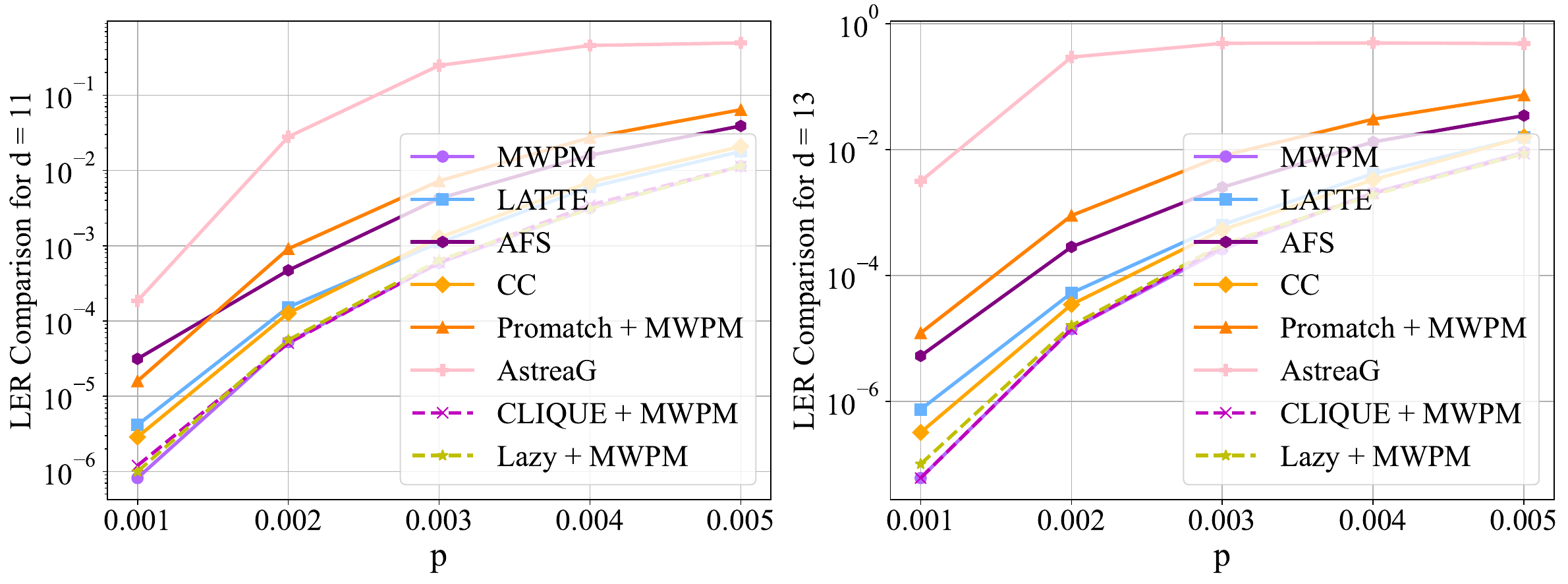}
\label{fig:memory experiment comp d = 11 and d = 13}
}
\caption{(a) Memory experiment and stability experiment. The memory experiment threshold of LATTE with NLDU is about $6.2 \times 10^{-3}$. The stability experiment threshold of LATTE with NLDU is about $7\times 10^{-3}$. Both thresholds are close to that of MWPM. (b) Long time memory experiment and multi-patch measurement experiment with increasing buffer size, $N_{block} = 10$ and $p = 0.003$ for simplicity. (c) Logical error rates in comparison with leading decoding architectures.}
\label{fig:accuracy experiment}
\end{center}
\end{figure}

The memory experiment~\cite{RyanAnderson2021RealizationOR,zhao2022realization,krinner2022realizing} focuses on the temporal fault tolerant ability, while the stability experiment~\cite{Gidney2022StabilityET} focuses on the spatial fault tolerant ability, both are critical during lattice surgery. We conduct memory and stability experiments across a wide range of code distances and physical error rates. LATTE achieves low logical error rates and a high threshold, comparable to MWPM, as shown in Figure~\ref{fig:memory experiment and stability experiment}.

\subsubsection{Multi-Block Decoding Experiment}
For long time quantum memory or multi-patch measurement, the accuracy of block decoding is guaranteed by the merging process in Figure~\ref{fig:decode and merge}. We conduct the long time quantum memory and multi-patch measurement experiments with increasing buffer size. Figure~\ref{fig:time and space buffer size experiment} demonstrates that as the buffer size $b \geq \lceil d / 2 \rceil$, the accuracy of asynchronous block decoding converges with that of the global MWPM decoding,
% not degrade in  comparing to monolithic decoding, 
consistent with the results reported in~\cite{skoric2023parallel,Tan2022ScalableSD,bombin2023modular}. 

\subsubsection{Accuracy Sensitivity Comparison}
The accuracy of LATTE is compared with various hardware implemented decoding works. Earliest FPGA accelerated decoder like LILLIPUT~\cite{das2022lilliput}, which only targets extremely small code distances like $d=3,5$, are not considered in our comparison as LATTE without NLDU can handle them easily. For NN decoders with high accuracy~\cite{wang2023transformer, bausch2024learning}, they can only operates at $d \leq 11$ while introducing extensive decoding latency ($>100 \mu s$ for $d = 5, round = 5$~\cite{bausch2024learning}) and non-scalable computational resources, making them unpractical for hardware deployment. As most of the leading decoders only focus on the $d\times d\times d$ decoding scenario and do not support arbitrary temporal-spatial decoding graphs, we evaluate the accuracy for code distance $d$ with $d$ measurement rounds. The comparison includes Lazy~\cite{delfosse2020hierarchical}, CLIQUE~\cite{ravi2023better}, Promatch~\cite{alavisamani2024promatch}, AstreaG~\cite{Vittal2023AstreaAQ}, AFS~\cite{das2022afs} and CC~\cite{barber2025real}. The results for $d=11$ and $d=13$ are shown in Figure~\ref{fig:memory experiment comp d = 11 and d = 13}.

Without the NLDU, LATTE has the same accuracy as MWPM~\cite{higgott2025sparse}, which is also consistent with Google and Fusion Blossom~\cite{acharya2025quantum, wu2023fusion}. The overall architecture of LATTE still remains high decoding accuracy among the hardware implemented decoders, even comparable to the CC decoder without predecoding. Note that both AFS and CC implement the UF~\cite{delfosse2021almost}, while CC has higher accuracy due to the ability for circuit-level noise decoding. Compared to latest predecoding works~\cite{alavisamani2024promatch, Vittal2023AstreaAQ}, LATTE demonstrates much higher accuracy at larger code distances under $p \sim 10^{-3}$ due to the capability to deal with complex scenarios in the circuit-level noise model, and decodes XZ simultaneously. Although the accuracy of~\cite{delfosse2020hierarchical, ravi2023better} is close to that of MWPM, this is due to their overly conservative strategy, which almost always returns failure and degrades into the MWPM decoder at high physical error rates or large code distances, resulting in little to no reduction in transmission bandwidth, as discussed in $\S$~\ref{sec:transmission bandwidth reduction} later.

\subsection{Transmission Bandwidth}
\label{sec:transmission bandwidth reduction}
\begin{figure}[htbp]
\centering
\includegraphics[width=\linewidth]{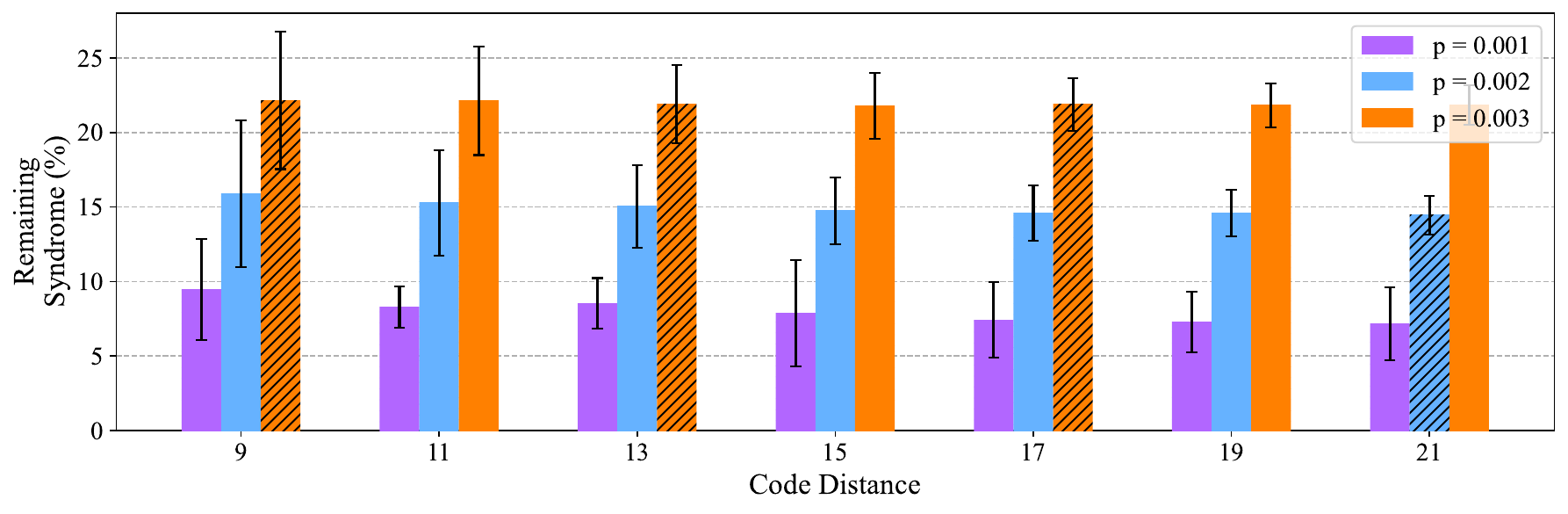}
\caption{Remaining syndrome ratio of LATTE after NLDU.}
\label{fig:bandwidth reduction}
\end{figure}

Under a physical error rate of $p=0.001$, corresponding to the current quantum hardware level, the NLDU in LATTE reduces bandwidth by over $90\%$, as shown in Figure~\ref{fig:bandwidth reduction}.

We also compare the remaining syndrome ratio with different predecoding methods in Table~\ref{table:bandwidth comparison}. The results highlight the effectiveness of our NLDU in reducing transmission bandwidth, significantly alleviating the data transmission burden. The syndrome reduction in~\cite{ravi2023better, delfosse2020hierarchical} approaches 0 as the code distance and physical error rate increase, making them hard to scale with current superconducting quantum computing systems~\cite{Acharya2022SuppressingQE, acharya2025quantum, gao2025establishing}.
\begin{table}[ht]
    \centering
    \small
    \caption{Comparison of Remaining Syndrome Ratio}
    \begin{tabular}{ccccccc}
        \toprule
        % \hline
        \textbf{d} & \textbf{p} & \textbf{Lazy} & \textbf{CLIQUE} & \textbf{Promatch} & \textbf{LATTE} \\
        \midrule
        % \hline
        11 & 0.001 & 73.96\% & 78.59\% & 35.77\%  & \textbf{8.28\%} \\
        % \hline
        13 & 0.001 & 88.90\% & 91.81\% & 34.26\%  & \textbf{8.53\%} 
        \\
        11 & 0.002 & 92.96\% & 96.20\% & 50.53\%  & \textbf{15.28\%} \\
        % \hline
        13 & 0.002 & 98.96\% & 99.59\% & 48.38\%  & \textbf{15.04\%} 
        \\
        11 & 0.003 & 98.56\% & 99.48\% & 60.20\%  & \textbf{22.11\%} \\
        % \hline
        13 & 0.003 & 99.92\% & 99.98\% & 59.40\%  & \textbf{21.91\%} 
        \\
        \bottomrule
        % \hline
    \end{tabular}
    \label{table:bandwidth comparison}
\end{table}

\begin{figure}[htbp]
    \centering
    % \subfigure[Decoding Latency of Arbitrary Rounds for $d = 13$]{
    %     % \hspace*{-0.075\linewidth}
    %     \includegraphics[width=0.85\linewidth]{Image/59.pdf}
    %     \label{fig:arbitrary rounds decoding latency}
    % }
    \subfigure[Single-Block Decoding Latency]{
        \includegraphics[width=0.85\linewidth]{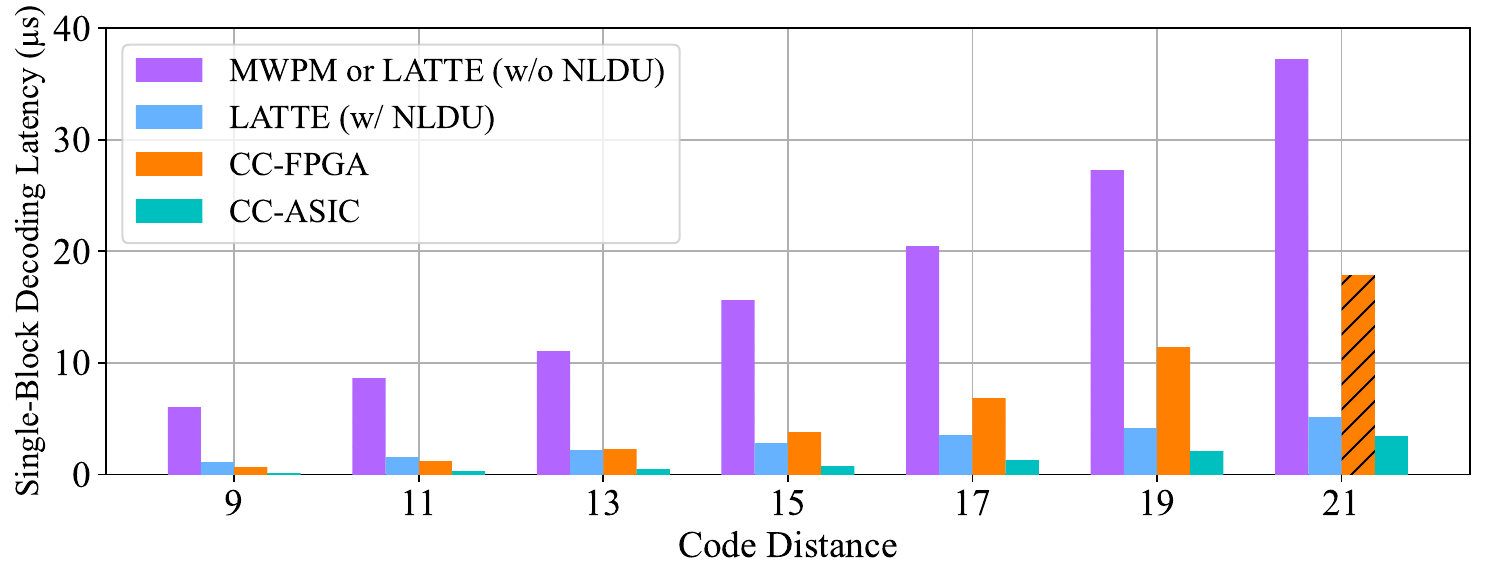}
        \label{fig:single block latency}
    }
    \subfigure[Constant Streaming Decoding Latency]{
        % \hspace*{-0.075\linewidth}
        \includegraphics[width=0.85\linewidth]{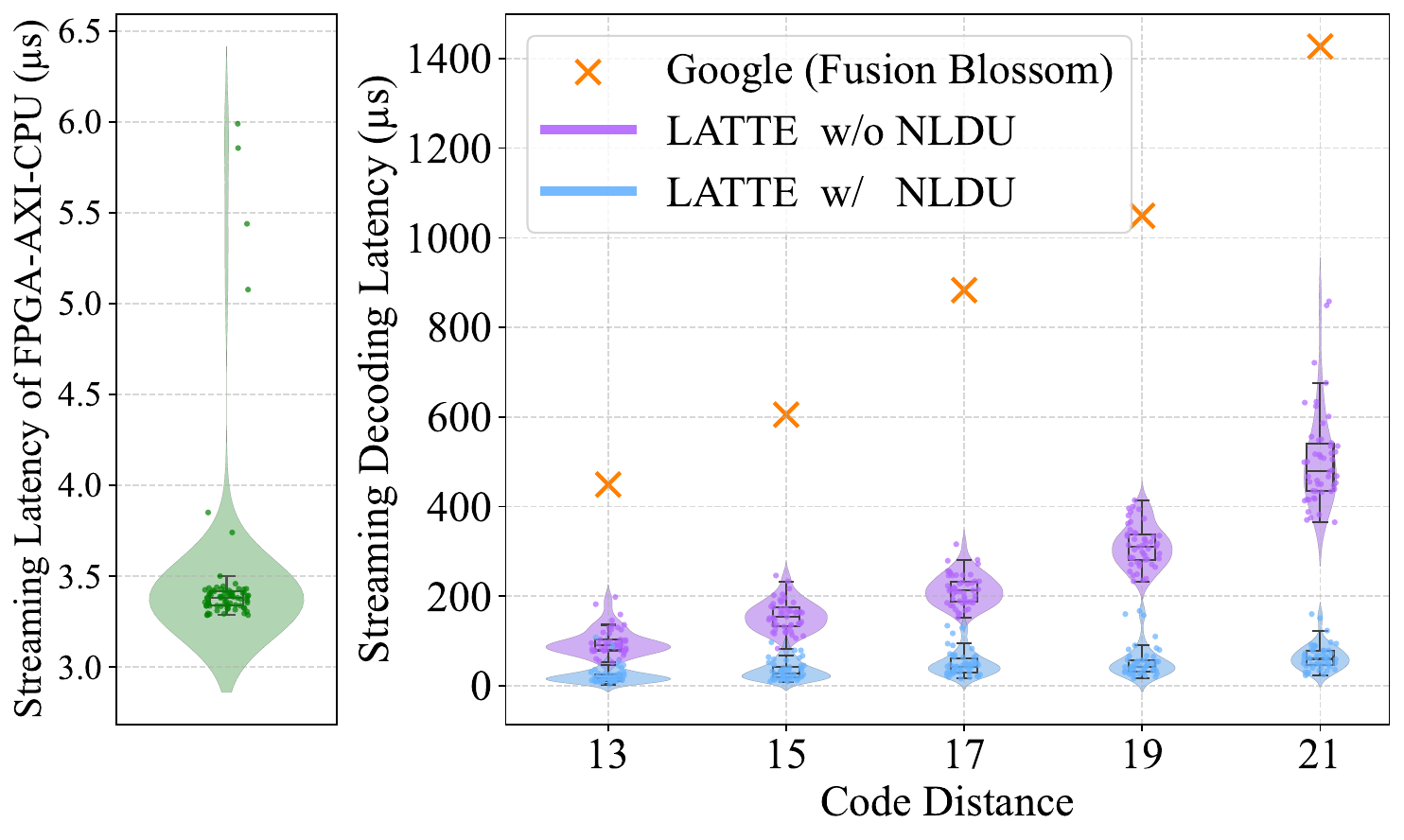}
        \label{fig:streaming decoding latency}
    }
    \subfigure[Multi-Patch Measurement Decoding Latency]{
        % \hspace*{-0.1\linewidth}
        \includegraphics[width=0.85\linewidth]{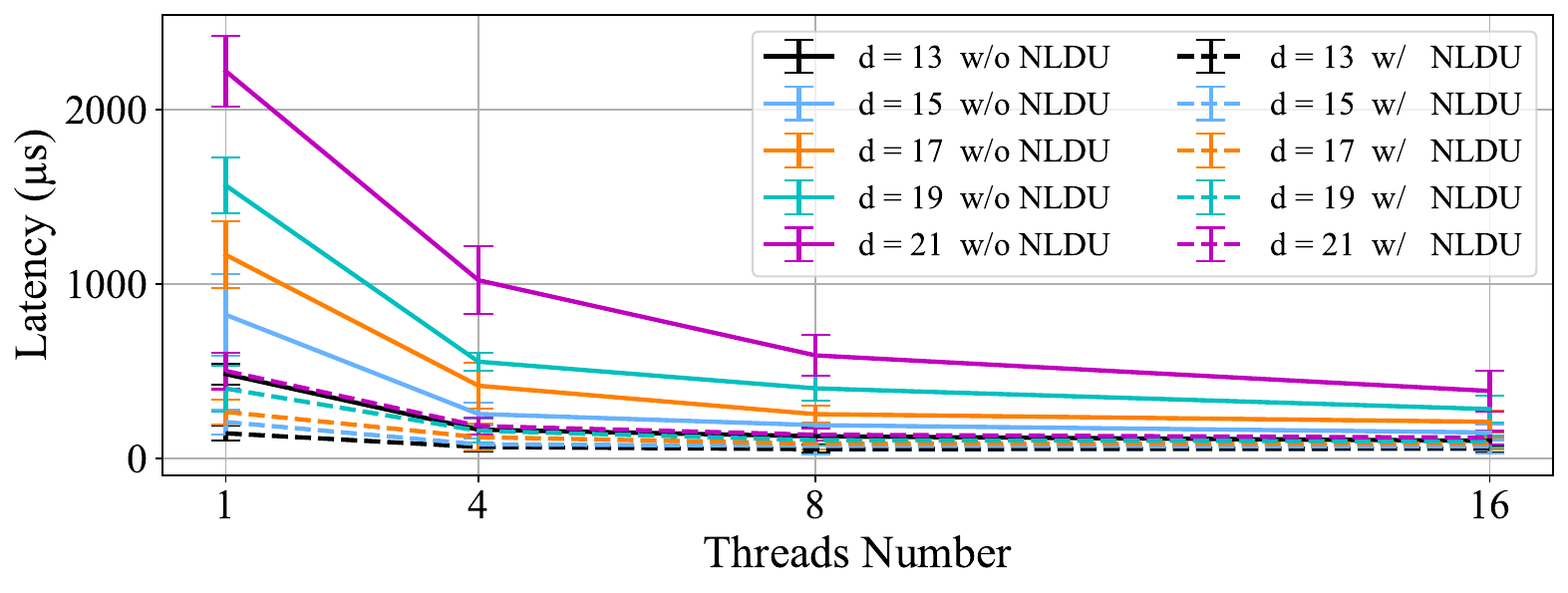}
        \label{fig:multi-space-block decoding latency}
    }
    \caption{(a) Single-block decoding latency. (b) Streaming decoding latency for arbitrary measurement rounds (up to $10^6$). (a) (b) are both evaluated by memory experiments under $p = 0.001$. (c) Multi-patch measurement decoding latency across different code distances and thread numbers, with $N_{patch} = 16, p = 0.001, b = \lceil d/2 \rceil$.}
\end{figure}

\subsection{Throughput}
\label{sec:throughput}
When evaluating the throughput of LATTE, we aim to answer two key questions: 
\begin{itemize}
    % \item Why is the streaming block decoding architecture required?
    \item Whether the architecture can support arbitrarily long streaming decoding without a syndrome backlog?
    \item Whether the system can efficiently parallelize decoding across different spatial regions?
\end{itemize}

% We first demonstrate that without streaming and parallel decoding, even for $d = 13$, the decoding throughput cannot be satisfied due to the increasing volume of the syndrome graph, in Figure~\ref{fig:arbitrary rounds decoding latency}. Therefore, the latency of Sparse Blossom and CC decoder grows almost linearly with the number of measurement rounds. While for LATTE, Google~\cite{acharya2025quantum} and Fusion Blossom~\cite{wu2023fusion}, they can maintain a constant latency regardless of the measurement rounds, thus enabling sufficient throughput for realistic FTQC process.

Through long time quantum memory experiment in Figure~\ref{fig:streaming decoding latency}, LATTE achieves constant and low latency across different code distances for arbitrary number of measurement rounds (tested up to $10^6$), ensuring that no backlog or memory overflow occurs. 

Through multi-patch measurement experiment in Figure~\ref{fig:multi-space-block decoding latency}, we observe that as the number of decoding threads approaches the number of logical patches, the latency reaches its constant optimum. Although the performance gain of spatial decoding becomes sub-linear for more threads due to overheads such as thread synchronization or memory contention, the overall trend still demonstrates that LATTE can leverage multi-threading effectively to achieve sufficient throughput.

\subsection{Latency}
LATTE achieves low decoding latency through the FPGA-CPU hybrid system, improving the fidelity of logical operations. We evaluate it using: (1) Single-block decoding (2) streaming decoding for arbitrary measurement rounds and (3) multi-patch measurement decoding. 

\subsubsection{Single-Block Decoding}
% like Promatch~\cite{alavisamani2024promatch} (only $d=11, 13$), LILLIPUT~\cite{das2022lilliput} (only $d = 3, 5$), CC~\cite{barber2025real} (with $d = 3 \sim 23$)
Existing decoder designs prior to Fusion Blossom~\cite{wu2023fusion} and Google’s real-time decoding architecture~\cite{acharya2025quantum} are typically benchmarked only on a $d \times d \times d$ syndrome block, without streaming decoding and arbitrary temporal-spatial volumes. To demonstrate LATTE’s performance in this context, we evaluate the decoding latency under code distance $d$ for $d$ syndrome measurement rounds at $p = 0.001$, as shown in Figure~\ref{fig:single block latency}. For a comprehensive comparison, we also include latency estimation (for $d$ rounds) from the CC decoder~\cite{barber2025real}, which is the only available hardware decoder providing experimental data under circuit-level noise with $p = 0.001$, aligning closely with our setup. Compared to the state-of-the-art MWPM implementation --- Sparse Blossom~\cite{higgott2025sparse}, LATTE without NLDU achieves the same latency, as no streaming or scheduling overhead is involved in this setting. For LATTE with the NLDU, the decoding can be accelerated by up to $6.4\times$ on average. Compared with the CC decoder, LATTE achieves even comparable latency with the ASIC implementation, demonstrating the efficiency of our FPGA-CPU hybrid system while providing greater scalability. Moreover, since our NLDU is implemented on the control hardware, it can also be integrated with latest FPGA or ASIC based decoders~\cite{liyanage2024fpga, barber2025real}, achieving even lower latency.

\subsubsection{Streaming Decoding}
In Figure~\ref{fig:streaming decoding latency}, LATTE demonstrates constant and significantly lower overall decoding latency in arbitrarily long memory experiment compared to Google’s real-time decoding system~\cite{acharya2025quantum} based on Fusion Blossom~\cite{wu2023fusion}. Under the same experimental setup, LATTE achieves a $16\times$-$20\times$ speedup, and still attains a $3\times$-$4\times$ speedup even without integrating the NLDU, attributed to the efficiency of BLDS's asynchronous decoding design.

\subsubsection{Multi-Patch Measurement Decoding}
For the multi-patch measurement experiment shown in Figure~\ref{fig:multi-space-block decoding latency}, LATTE achieves a near-linear speedup as the number of threads increases compared to global MWPM (single thread), and delivers an average $3\times$ speedup over LATTE without the NLDU, demonstrating the ability to decode parallel logical operations in lattice surgery.

\subsection{Hardware Resource Utilization}
\label{sec:hardware estimation}
In the end, we report hardware resources required of LATTE, including the CPU and FPGA.

\subsubsection{CPU Resource Usage}
Unlike the streaming decoding design in~\cite{wu2023fusion, acharya2025quantum}, where decoding and fusion are interdependent, the merging overhead during asynchronous block decoding in LATTE is negligible. We simply choose a fixed number of $2$ merging threads to handle it. Table~\ref{table:time decoding latency} shows that a near-optimal number of decoding threads can achieve fixed latency for arbitrary measurement rounds. We demonstrate LATTE's scalability by using much less computational resources to achieve sufficient throughput and lower latency, compared with Google (Fusion Blossom)'s implementation.

\begin{table}[h!]
    \centering
    \small
    \caption{Minimum Threads Required, $p = 0.001, b = \lceil d/2 \rceil$, with Varying NLDU Configurations (NC)}
    \begin{tabular}{c|c|c|c|c}
    \toprule
    \textbf{d} & \textbf{Google} & \textbf{Ours (w/o NLDU)} & \textbf{Ours (w/ NLDU)} & \textbf{\#NC$^1$}\\ 
    \midrule
    13 & 16 & 6  & \textbf{2} & 1\\ 
    
    15 & 20 & 8  & \textbf{2} & 1\\ 
    
    17 & 24 & 12 & \textbf{2} & 4\\ 
    
    19 & 30 & 16 & \textbf{4} & 4\\ 
    
    21 & 32 & 24 & \textbf{6} & 4\\ 
    
    \bottomrule
    \end{tabular}
    \label{table:time decoding latency}
    \begin{flushleft}
    \footnotesize{$^1$ Each NLDU controls a $N^2 = 9{\times}9$ region with a receptive field $R = 7$. Consequently, one NLDU covers up to $15{\times}15$ patch size. $d \leq 15$ fit in 1 NLDU; $d \geq 17$ require 4 NLDUs.}
    \end{flushleft}
\end{table}

\subsubsection{FPGA Resource Usage}
% Due to the configurable IP design, we also derive empirical formulas in Equations~\eqref{formula:all_estimations} to estimate the $DSP, LUT, REG$ usage and latency $LTC$ based on $N, {K_i}$, INT$M$, $P_i$, and the clock frequency $f$.
% Based on the configurable IP design, 
The NLDU design based on $N=9$, $K_{1:3}=7$ (the same as Table~\ref{tab:model arch}) and $P_1, P_2, P_3=52, 33, 27$ is implemented on \emph{Xilinx Alveo U200} boards. The number of NLDUs required for a distributed deployement is shown in Table~\ref{table:time decoding latency}. The actual resources of a single NLDU-FPGA board and inference latency of each stage are listed in Table~\ref{table:FPAG Quantization Overhead},~\ref{table:FPAG Inference Latency}. Currently, most multipliers and adders are mapped to LUTs due to the INT8 quantization. However, mapping more of these operations to DSP slices can further reduce latency, area usage, and power consumption. Latency of each stage is fixed and less than $1 \mu s$, allowing for pipelined inference without backlog. The end-to-end streaming latency from FPGA to CPU is also tested in Figure~\ref{fig:streaming decoding latency}, which is close to the estimation in Equation~\eqref{formula:Hardware estimation LATENCY} with some deviation due to the boundary case in the last measurement round (does not require additional idle in the 3-stage streaming pipeline), and the communication latency.

% \begin{table}[h!]
% \centering
% \small
% \caption{FPGA Resources Usage}
% \begin{tabular}{|c|c|c|c|c|}
% \hline
% \textbf{Quant} & \textbf{DSP} & \textbf{LUT(s)} & \textbf{REG(s)} & \textbf{On-chip SRAM} \\ 
% \hline
% \hline
% \multirow{2}{*}{\textbf{INT8}} & 5264 & 178339  & 478632 & 6.07KB \\
% \cline{2-5}
% & 42.84\% & 10.08\% & 13.52\% & 0.011\% \\
% \hline
% \end{tabular}
% \label{table:FPAG Quantization Overhead}
% \end{table}

\begin{table}[h!]
\centering
\small
\caption{FPGA Resources Utilization}
\begin{tabular}{|c|c|c|c|c|c|}
\hline
\textbf{Quant} & \textbf{DSP} & \textbf{LUT(s)} & \textbf{REG(s)} & \textbf{BRAM} & \textbf{URAM} \\ 
\hline
\hline
\multirow{2}{*}{\textbf{INT8}} & 16 & 270338  & 393383 & 451.5 & 20\\
\cline{2-6}
& 0.23\% & 22.87\% & 16.64\% & 20.9\% & 2.08\%\\
\hline
\end{tabular}
\label{table:FPAG Quantization Overhead}
\end{table}

% \begin{scriptsize}
\begin{table}[h!]
    \centering
    \small
    \caption{Inference Latency of Each Stage}
    \begin{tabular}{cccc}
    \toprule
    \textbf{Clock Speed} & $\mathbf{ST_1}$ & $\mathbf{ST_2}$ & $\mathbf{ST_3}$ \\ 
    \midrule
    300 MHz & 433 ns & 433 ns & 346 ns \\ 
    \bottomrule
    \end{tabular}
    \label{table:FPAG Inference Latency}
\end{table}
% \end{scriptsize}

\section{Summary and Outlook}
In this work, we present LATTE --- a hierarchical, fully streaming and distributed decoding architecture designed for efficient deployment with modest resources. The architecture meets all decoding requirements at scale, serving as a core component of a high-performance, fault-tolerant quantum computer. Its FPGA-CPU hybrid setup not only leverages current quantum control electronics for seamless deployment but also enables unmatched scalability. LATTE's enhanced scalability further enables operation with large-scale quantum hardware beyond conventional assumptions --- supporting error rates at $p \sim 10^{-3}$, where all current quantum devices operate. This unique capability positions our system as a unified architecture: immediately deployable on today's hardware while remaining scalable for future systems.

We comprehensively evaluated LATTE’s standalone performance and its integrated behavior with control electronics. These results establish the foundation for the next critical research phase: deployment on large-scale quantum processors for full end-to-end system validation. This step will provide key insights into the scalability and future development of the system. Furthermore, while LATTE achieves better overall performance with only slight (often imperceptible) sacrifices in global decoding accuracy, further enhancing accuracy remains an important goal for long-term progress.

\section*{ACKNOWLEDGMENTS}
The work is supported by National Key Research and Development
Program of China (Grant No.\ 2023YFA1009403), National Natural Science
Foundation of China (Grant No.\ 12347104), Beijing Natural Science Foundation
(Grant No.\ Z220002), Zhongguancun Laboratory, and Tsinghua University.

%%%%%%% -- PAPER CONTENT ENDS -- %%%%%%%%

%%%%%%%%% -- BIB STYLE AND FILE -- %%%%%%%%
\clearpage
\bibliographystyle{plain}
\bibliography{reference}
%%%%%%%%%%%%%%%%%%%%%%%%%%%%%%%%%%%%

\end{document}